\documentclass[prd,singlecolumn,superscriptaddress,nofootinbib,notitlepage]{revtex4-1}

\usepackage{epsfig,amsmath,amssymb,slashed}
\usepackage{graphicx}
\usepackage{color}
\usepackage{siunitx}
\usepackage{color}
\usepackage{bm}
\usepackage{subfig}
\usepackage[utf8]{inputenc}
\usepackage[english]{babel}
\usepackage[T1]{fontenc}
\usepackage{amsfonts}
\usepackage{amsthm}
\usepackage{hyperref}
\usepackage{wasysym}
\usepackage{comment}
\usepackage[dvipsnames]{xcolor}

\newcommand{\nn}{{\newline}}

\newcommand{\ped}{{\cal E}}
\newcommand{\pres}{{\cal P}}
\newcommand{\taueq}{\tau_{\rm eq}}
\newcommand{\feq}{f_{\rm eq}}
\newcommand{\pedeq}{{\cal E}_{\rm eq}}
\newcommand{\preseq}{{\cal P}_{\rm eq}}
\newcommand{\Teq}{T_{\rm eq}}
\newcommand{\rhoeq}{\rho_{\rm eq}}

\newcommand{\fplus}{f^{+}}

\newcommand{\fpluseq}{f^{+}_{\rm eq}}

\newcommand{\fpm}{f^{\pm}}

\newcommand{\Jeq}{J_{\rm eq}}
\newcommand{\Tk}{T_{0}}

\newcommand{\fp}{f^{+}}
\newcommand{\fm}{f^{-}}

\newcommand{\fn}{f_{\rm 0}}
\newcommand{\fnp}{f_{\rm 0}^+}
\newcommand{\fnm}{f_{\rm 0}^-} 
\newcommand{\fnpm}{f_{\rm 0}^{\pm}} 
\newcommand{\fnmp}{f_{\rm 0}^{\mp}} 
\newcommand{\fnq}{f_{\rm 0}^{q}} 
\newcommand{\fnqbar}{f_{\rm 0}^{{\bar q}}} 
\newcommand{\mq}{{M}} 

\newcommand{\bg}{{\boldsymbol{g}}} 
\newcommand{\bq}{{\boldsymbol{q}}}

\newcommand{\deltaB}{\delta\! B}

\newcommand{\Tkin}{T_{\rm kin}}
\newcommand{\Tkinzero}{T_{0}}

\newcommand{\Tkindot}{{\dot T}_{\rm kin}}
\newcommand{\Tkinzerodot}{{\dot T}_{0}}

\begin{document}

%*************************************************************************************
\title{%Transport coefficients in quasi-particle relativistic second-order hydrodynamics at finite chemical potential\textcolor{red}{?}}
Quasiparticle second-order dissipative hydrodynamics at finite chemical potential} 
%*************************************************************************************
\author{Asaad Daher}
\email{asaad.daher@ifj.edu.pl}
\affiliation{Institute of Nuclear Physics Polish Academy of Sciences, PL-31-342 Krakow, Poland}
\author{Leonardo Tinti}
\email{dr.leonardo.tinti@gmail.com}
\affiliation{Institute of Physics, Jan Kochanowski University, ul. Uniwersytecka 7, 25-406, Kielce, Poland}
\author{Amaresh Jaiswal}
\email{a.jaiswal@niser.ac.in}
\affiliation{School of Physical Sciences, National Institute of Science Education and Research, An OCC of Homi Bhabha National Institute, Jatni-752050, Odisha, India}
\author{Radoslaw Ryblewski}
\email{radoslaw.ryblewski@ifj.edu.pl}
\affiliation{Institute of Nuclear Physics Polish Academy of Sciences, PL-31-342 Krakow, Poland}

\date{\today}

%-------------------------------------------------------------------
\begin{abstract}

 %We consider two quasi-particle's species, charged ones, which contribute both the energy-momentum and baryon current, and uncharged, which contributes only to the energy and momentum flux. We consider also two running masses, dependent on the temperature and chemical potential, one for the charged and one for the uncharged quasi-particle species. We devise a thermodynamically consistent framework to formulate the dynamics and extract the second-order viscous hydrodynamic equations as an application of the framework. 
 
We extend the derivation of second-order relativistic viscous hydrodynamics to incorporate the effects of baryon current, a non-vanishing chemical potential, and a realistic equation of state.
Starting from a microscopic quantum theory, we employ a quasiparticle approximation to describe the evolution of hydrodynamic degrees of freedom and establish its connection to the Wigner formalism. Using methods from relativistic kinetic theory, we perform a second-order expansion to derive a closed set of equations for the components of the stress-energy tensor and the baryon current. The resulting transport coefficients, which depend on the equation of state, are obtained through a unified prescription that ensures thermodynamic consistency.

%We extend the derivation of second-order relativistic viscous hydrodynamics to include the effects of the baryon current, a non-vanishing chemical potential, and a realistic equations of state.

%We first discuss the microscopic quantum theory, a quasi-particle approximation to describe the evolution of the hydrodynamic degrees of freedom, and its link to the Wigner formalism. Extending the methods used in relativistic kinetic theory, we perform a second order approximation to obtain closed set of equations for the components of the stress-energy tensor and the baryon current. The obtained transport coefficients are sensitive to the equation of state and ar obtained from a singe prescription without thermodynamic inconsistency.

%We present the derivation of relativistic second-order dissipative hydrodynamics by considering an effective Boltzmann equation for a system composed of a non-interacting mixture of three quasi-particle species: quarks, anti-quarks, and gluons. Our approach incorporates temperature and chemical potential-dependent masses for these quasiparticles and establishes a thermodynamically consistent framework for formulating second-order evolution equations pertaining to corrections in shear and bulk viscous pressure as well as baryon charge diffusion.
%
\end{abstract}
%-------------------------------------------------------------------

\maketitle

%*************************************************************************************
\section{Introduction}

Relativistic fluid dynamics has been extensively applied to analyze the space-time evolution of strongly interacting, hot, and dense matter produced in ultra-relativistic heavy-ion collisions at the Relativistic Heavy-Ion Collider (RHIC) and the Large Hadron Collider (LHC)~\cite{Busza:2018rrf, Huovinen:2006jp, Romatschke:2009im, Heinz:2013th, Gale:2013da, DerradideSouza:2015kpt, Jaiswal:2016hex, lisa2021}. The early success of pioneering studies using non-dissipative fluid dynamics to interpret experimental data from RHIC introduced the concept of a perfect fluid state, commonly referred to as the quark-gluon plasma (QGP)~\cite{Shuryak:2003xe}. However, these findings also spurred investigations into the rapid thermalization of the QGP and the limits of applicability for relativistic hydrodynamics~\cite{Heinz:2001xi}. To account for the omnipresent dissipative effects in nature and to enhance the precision of flow measurements at the LHC, the incorporation of viscous effects into fluid-dynamic models became imperative~\cite{Danielewicz:1984ww, Policastro:2001yc, Kovtun:2004de}. This need has driven significant progress in the development of relativistic viscous hydrodynamics~\cite{Muronga:2003ta, Baier:2006um,York:2008rr, El:2009vj, Denicol:2010xn, Denicol:2012cn, Denicol:2012es, Jaiswal:2012qm, Bhalerao:2013aha, Rocha:2023ilf}, motivated by the growing recognition of the importance of viscosity and its role in heavy-ion collisions. Consequently, substantial research has focused on extracting the thermodynamic and transport properties of the QGP medium, such as shear viscosity ($\eta$), bulk viscosity ($\zeta$), and charge diffusion ($\kappa$)~\cite{Romatschke:2007mq, Bozek:2009dw, Song:2010mg, Alver:2010dn, Schenke:2011bn, Gale:2012rq, Noronha-Hostler:2013gga, Gardim:2014tya, Ryu:2015vwa, Niemi:2015qia}. Recent studies have further advanced these investigations by incorporating a spin current, providing deeper insights into the transport properties of relativistic spin hydrodynamics~\cite{Florkowski:2018fap,Becattini:2023ouz,Huang:2024ffg}.

The success of viscous fluid dynamics in explaining the various collective phenomena observed in high-energy heavy-ion collisions is often attributed to the near-equilibrium nature of the QGP. This assumption is foundational to the formulation of relativistic hydrodynamics, which is typically constructed as an expansion around local thermodynamic equilibrium in terms of thermodynamic gradients. The zeroth-order term corresponds to non-dissipative hydrodynamics. However, the first-order relativistic Navier-Stokes theory~\cite{Eckart:1940zz}, which involves parabolic differential equations, suffers from acausality and instability issues~\cite{Hiscock:1983zz, Denicol:2008ha}. To address these challenges, the second-order Israel-Stewart theory was developed~\cite{Israel:1976tn, Israel:1979wp}. This formulation incorporates hyperbolic equations that restore causality, although stability is not always guaranteed~\cite{Van:2007pw, Pu:2009fj}. More recently, efforts have been made to develop formulations of causal and stable relativistic hydrodynamics, even at first order in gradients~\cite{Bemfica:2017wps, Kovtun:2019hdm, Abbasi:2022rum}. An analogous discussion emphasizing the necessity of a second-order spin hydrodynamic theory~\cite{Weickgenannt:2022zxs,Biswas:2023qsw,Tiwari:2024trl,She:2024rnx} has also gained traction, driven by concerns over stability and causality~\cite{Daher:2022wzf,Ren:2024pur}. 

To capture the collective behavior of the quark-gluon plasma (QGP) using fluid dynamics, it is essential to incorporate its properties through transport coefficients and the equation of state.  Ideally, these properties, along with the regime of applicability of hydrodynamics, should be determined directly from experimental measurements. However, this is exceedingly challenging in the context of heavy-ion collisions because of the lack of controlled methods for probing the medium and extracting this information.  
Consequently, alternative methodologies are often employed. 

A comprehensive understanding of the properties of nuclear matter created in relativistic heavy-ion collisions should emerge from Quantum Chromodynamics (QCD), the fundamental theory of strong interactions. One strategy to achieve this goal is to integrate the results from ab initio calculations of thermodynamic and transport properties, obtained by lattice QCD (lQCD) calculations, into the hydrodynamic framework~\cite{Fodor:2009ax, Bazavov:2009zn, Borsanyi:2010cj, HotQCD:2014kol, Borsanyi:2013bia}. Unlike the non-relativistic Navier-Stokes theory, however, the equations governing relativistic dissipative hydrodynamics are not universally defined and depend strongly on the specific microscopic theory used in their derivation~\cite{Florkowski:2013lza, Florkowski:2013lya, Florkowski:2014sfa, Jaiswal:2014isa, Denicol:2014mca}. Moreover, in the phenomenologically relevant regime, lQCD calculations of transport properties still face significant uncertainties~\cite{Meyer:2007ic, Meyer:2007dy}.

Given the aforementioned challenges, a simplified microscopic theory, like kinetic theory, is often used to derive hydrodynamic evolution equations. Further, realistic properties of strongly interacting matter are incorporated by employing the equation of state and transport coefficients obtained from lattice QCD calculations. However, this approach inadvertently fixes the parameters of the microscopic theory, thereby introducing effective interactions that should be accounted for during the derivation of the hydrodynamic evolution equations. For instance, in the simplest case of relativistic kinetic theory for a single particle species, the particle mass determines the equation of state. This straightforward formalism for a constant particle mass does not lead to an agreement with the temperature dependence of energy density and pressure provided by lQCD. To achieve agreement with the lQCD results, temperature-dependent particle masses may be considered, leading to a non-ideal equation of state within a thermodynamically consistent framework~\cite{Gorenstein:1995vm, Romatschke:2011qp, Sasaki:2008fg, Bluhm:2010qf}. 

In such approaches, thermodynamic consistency is achieved by adding an extra contribution to the energy-momentum tensor, which can be interpreted as a physically significant mean-field interaction in the Boltzmann equation. While this procedure is equivalent to introducing the concept of interacting quasiparticles in the microscopic theory, it is important to note that they can only be considered as actual quasiparticle excitations of the fundamental theory in specific limits~\cite{Romatschke:2011qp}. At present, the construction of an effective quasiparticle kinetic theory that can accurately reproduce any thermodynamically consistent equation of state at finite chemical potential remains an open task. Consequently, the derivation of causal hydrodynamic evolution equations by coarse-graining such a theory has not yet been accomplished. In this paper, we aim to address this gap by deriving evolution equations for second-order dissipative hydrodynamics for a system composed of charged quasiparticles (as well as their antiparticles) and uncharged ones. These quasiparticles cannot correspond to known particles since, depending on the stage of the expansion, are expected to represent either (dressed) quarks and gluon, or mesons and Hadrons. 

We start with a general discussion in Section~\ref{Sec:quasiparticles}. We discuss the general properties of a system, and highlight some exact formulas that will be used to simplify the calculations. More importantly we introduce the quasiparticles plus bag model aimed at describing the hydrodynamic degrees of freedom. In section~\ref{sec:therm} we discuss the global equilibrium properties, and how to insert a generic equation of state in the quasiparticles model. Section~\ref{sec:dynamics} is dedicated to the dynamics of the bag term. In Section~\ref{sec:second_order} the equations of second-order hydrodynamics and the transport coefficients are obtained from the model. The conlusions and final remarks in Section~\ref{sec:conclusions}. 

Unless otherwise stated we use natural units where $\hbar = c = k_B=1$. We adopt the Einstein convention of automatically summing over repeated upper and lower indices, and we represent the contraction (scalar product) between four-vectors with a dot: $v^\mu w_\mu = v\cdot w$. Likewise, the Frobenius product reads $v^{\mu\nu} w_{\mu\nu} = v : w$. The ``mostly minus'' convention for the Minkowski metric is used, i.e. $g^{\mu\nu}={\rm diag}(1,-1,-1,-1)$, as well as the convention $\varepsilon^{0123}=1$ for the four-dimensional Levi-Civita symbol. Round parentheses around groups of Lorentz indices indicate symmetrization, e.g. $A^{(\mu} B^{\nu)} = \frac{1}{2!}(A^\mu B^\nu + A^\nu B^\mu)$.
The operator $\Delta^{\mu\nu}{\,=\,}g^{\mu\nu}{-}u^\mu u^\nu$ projects on the spatial coordinates in the comoving frame. Angular brackets around a tensor index indicate its spatial components in the LRF obtained by contracting the four-index with the spatial projector $\Delta^\mu_\nu$: $A^{\langle\mu\rangle}\equiv\Delta^\mu_\nu\, A^\nu$. The comoving time derivative is denoted by a dot ($\dot{A} \equiv u^\mu\partial_\mu A$ for any quantity $A$), and $\nabla^\mu\equiv\partial^{\langle\mu\rangle}\equiv\Delta^{\mu\nu}\partial_\nu$. The expansion rate is defined as $\theta=\partial_{\mu}u^{\mu}$.

%*************************** 
\section{Quasiparticles from quantum field theory}\label{Sec:quasiparticles}
%***************************
%
Relativistic hydrodynamics, much like its nonrelativistic counterpart, is based on the assumption that the conservation laws are sufficient to describe the system. In other words, the hydrodynamic degrees of freedom can be determined by solving a closed set of equations. Starting from the local energy, momentum, and baryon number conservation
\begin{equation}\label{conservation}
0=\partial_\mu J^\mu, \qquad 0=\partial_\mu T^{\mu\nu}.
\end{equation}
Identifying the components of the baryon current $J^\mu$ and stress-energy tensor $T^{\mu\nu}$ with the hydrodynamic degrees of freedom, one has only five independent equations for fourteen (assuming a symmetric $T^{\mu\nu}$, to which we limit our discussion in the current work) degrees of freedom.

From a fundamental point of view, the tensors $J^\mu$ and $T^{\mu\nu}$ are the expectation values of the corresponding operators in quantum field theory, with respect to density matrix $\hat\rho$
\begin{equation}\label{averages}
    T^{\mu\nu} = {\rm tr}\left( \hat \rho \, \widehat T^{\mu\nu} \right), \qquad J^{\mu} = {\rm tr}\left( \hat \rho \, \widehat J^\mu \right).
\end{equation}
In principle, from the action of the quantum system, one can write the explicit expression of the quantum operators $\widehat T^{\mu\nu}$ and $\widehat J^\mu$ in terms of the fundamental fields. Having an exact solution of the evolution of the fields and an exact expression of the (pure or mixed) state $\hat \rho$, one could perform the trace and have the values of $T^{\mu\nu}(x)$ and $J^\mu(x)$ at any spacetime point $x$.

Needless to say, such a daunting task is practically impossible, not only for the standard model or just the QCD sector but for almost all interacting field theories. Hence, there is a need for a phenomenological approximation. The assumption that the evolution of the hydrodynamic degrees of freedom can be adequately approximated with a closed set of equations is an assumption about the state as much as the fields. The evolution of $T^{\mu\nu}$ and $J^\mu$ stems from the evolution of the fields, which depends, in turn, on different combinations of the quantum fields compared to the ones that constitute $\widehat T^{\mu\nu}$ and $\widehat J^\mu$. Determining the extent to which the evolution of non-hydrodynamic degrees of freedom is influenced by the values of hydrodynamic degrees of freedom, rather than the reverse, is a difficult challenge to predict.
%How much the evolution of such non-hydrodynamic degrees of freedom is dominated by the values of the hydrodynamic degrees of freedom rather than the other way around is rather complicated to predict. 
The usual approach is to formulate some assumptions and verify with experiments if the resulting approximation is reliable.

Before continuing the discussion over hydrodynamics and its derivations, it is convenient to introduce a common subdivision of the hydrodynamic degrees of freedom. In most cases, the collective behavior of matter in heavy-ion collisions is referred to in terms of the flow (four-velocity), energy, and baryon number density, or, equivalently, the effective temperature and baryon chemical potential. There are other possible definitions, but in this work, we will use only Landau's prescription to define the four-velocity, that is, the time-like unitary eigenvector of $T^{\mu\nu}$
\begin{equation}\label{u_Landau}
T^{\mu\nu} u_\nu = u_\nu T^{\nu\mu} = \ped u^\mu, \qquad u\cdot u =1,
\end{equation}
the eigenvalue $\ped$ being called the proper energy density~\footnote{That is, the energy density in a reference frame comoving with the four-velocity $u$ at that point (as opposed to the energy density in the laboratory frame $T^{00}$).}. Making projections of the baryon current along the four-velocity and orthogonal to it, one recognizes the proper baryon density $\rho$ and the diffusion current $\nu^\mu$
\begin{equation}\label{J_dec}
    \rho = J\cdot u, \qquad \nu^\mu = \Delta^\mu_\alpha J^\alpha = \left( \vphantom{\frac{}{}} g^{\mu}_{\;\alpha} - u^\mu u_\alpha \right) J^\alpha, \qquad \Rightarrow \quad J^\mu = \rho \, u^\mu + \nu^\mu,
\end{equation}
where we introduced the projector onto the space orthogonal to $u$, $\Delta^\mu_\alpha$.

Similarly one can decompose the stress-energy tensor. Employing the definition in Eq.~\eqref{u_Landau}, only the proper energy part can be proportional to the four-velocity
\begin{equation}
    T^{\mu\nu} = \left( \vphantom{\frac{}{}} u^\mu u_\rho + \Delta^\mu_{~\rho} \right) T^{\rho\sigma} \left( \vphantom{\frac{}{}} u^\nu u_\sigma + \Delta^\nu_{~\sigma} \right) = \ped u^\mu u^\nu + \pres^{\mu\nu}.
\end{equation}
In the above equation, the pressure tensor $\pres^{\mu\nu}$ constitutes the only remaining degrees of freedom, since, by the definition of the Landau four-velocity itself~(\ref{u_Landau}), the mixed term $u\cdot T \cdot \Delta =0$ vanishes. Its usual decomposition involves separation of the trace part which is proportional to the projector $\Delta^{\mu\nu}$ itself, and the traceless part which is identified with the shear correction $\pi^{\mu\nu}$, i.e.,
\begin{equation}
    \begin{split}
        \pres^{\mu\nu} &= \Delta^\mu_{~\alpha} \Delta^\nu_{~\beta} \, T^{\alpha\beta} = \frac{1}{3}\left( \Delta_{\alpha\beta} T^{\alpha\beta} \right)\Delta^{\mu\nu} + \left(\Delta^\mu_{~\alpha} \Delta^\nu_{~\beta} -\frac{1}{3} \Delta_{\alpha\beta} \Delta^{\mu\nu} \right) T^{\alpha\beta} =-\left( \vphantom{\frac{}{}} \pres +\Pi\right)\Delta^{\mu\nu} +\pi^{\mu\nu}.
        \label{separ}
    \end{split}
\end{equation}
The separation of the trace part in the hydrostatic pressure $\pres$ and the bulk pressure correction $\Pi$ in Eq.~(\ref{separ}) is not geometric, but is given by the Landau matching and the equation of state. Specifically, as long as the relation at global equilibrium between the energy density $\pedeq(T,\mu)$ and baryon density $ \rhoeq(T,\mu)$ and the intensive parameters $T$ and $\mu$ is invertible, i.e., the Jacobian is non-vanishing %(and it is usually the case)
\begin{equation}\label{Jacobian}
    \frac{\partial \pedeq(T,\mu)}{\partial T} \frac{\partial \rhoeq(T,\mu)}{\partial \mu}  - \frac{\partial \pedeq(T,\mu)}{\partial \mu}  \frac{\partial \rhoeq(T,\mu)}{\partial T}  \neq 0,
\end{equation}
it is possible to define the effective temperature and effective baryon chemical potential from the proper energy and baryon density, using the inverse formulas applied to the physical densities
\begin{equation}\label{eff_T-mu}
    v=\left( \begin{matrix}
        \ped \\ \rho
    \end{matrix} \right) = {\sf f}(T,\mu) = \left( \begin{matrix}
        \pedeq(T,\mu) \\ \rhoeq(T,\mu)
    \end{matrix} \right), \qquad \Rightarrow \qquad \left( \begin{matrix}
        T(\ped,\rho) \\ \mu(\ped,\rho)
    \end{matrix} \right) = {\sf f}^{-1}(v).
\end{equation}
The hydrostatic pressure is defined as a function of the proper energy and baryon density, given by the equilibrium relations
\begin{equation}\label{L_match}
    \pres(\ped, \rho) = \preseq(T(\ped,\rho),\mu(\ped,\rho)).
\end{equation}
In other words, it is the pressure that a system would have at global equilibrium if the temperature and baryon chemical potential coincides with the effective temperature and baryon chemical potential of the local fluid cell. The bulk pressure correction is the remainder of the scalar (trace) part of the physical pressure tensor. 

Summing up, the decomposition of the stress-energy tensor reads
\begin{equation}\label{T_dec}
    T^{\mu\nu} = \ped \, u^\mu u^\nu -\left( \vphantom{\frac{}{}} \pres + \Pi \right)\Delta^{\mu\nu} + \pi^{\mu\nu}.
\end{equation}
The procedure of decomposing the physical tensors in the geometric sectors dictated by the four-velocity, obtaining the effective intensive parameters, inverting the relations at global equilibrium, and highlighting the part of the isotropic pressure that can be estimated from such global equilibrium relations is called Landau matching. It is also important to note that, so far, the treatment is rather general. No assumptions have been made about the dynamics. This decomposition is particularly useful for fluid systems, where densities and the four-velocity play an essential role, but it can be applied in any situation. It is convenient given the approximations we will make later since they directly use the effective temperature and baryon chemical potential. In any case, it is possible to show a very useful, exact result. Making use of the definition of the effective temperature $T$ and baryon chemical potential $\mu$~(\ref{eff_T-mu}), the matching~(\ref{L_match}) and the conservation equations~(\ref{conservation}), one can write exactly the comoving derivatives $\dot T$ and $\dot \mu$ of the effective parameters in terms of the hydrodynamic variables. 

Starting from expressing $\ped$ and $\rho$ as functions of the effective parameters~\footnote{Thanks to the definition of the effective parameters~(\ref{eff_T-mu}), as the ones that would give the physical densities if used in the equilibrium formulas.}
\begin{equation}\label{def_T_mu}
    \ped(x) = \pedeq(T(x),\mu(x)), \qquad \rho = \rhoeq(T(x),\mu(x)),
\end{equation}
and subsequently obtaining their exact co-moving derivatives
\begin{equation}\label{dot_T-mu}
\begin{split}
    \dot{ \ped} &=  (u\cdot\partial) \, \ped = (u\cdot\partial) \, \pedeq(T,\mu) =\frac{\partial \pedeq}{\partial T} \, \dot T + \frac{\partial \pedeq }{\partial \mu} \, \dot \mu,\\
    \dot{ \rho} &=  (u\cdot\partial) \, \rho = (u\cdot\partial) \, \rhoeq(T,\mu) =\frac{\partial \rhoeq}{\partial T} \, \dot T + \frac{\partial \rhoeq }{\partial \mu} \, \dot \mu,
\end{split}
\end{equation}
it is possible to write the left-hand side in terms of the hydrodynamic degrees of freedom only. Indeed, taking the projection of the local four-momentum conservation and the baryon number conservation from~(\ref{conservation}) along $u^\mu$, and using the decompositions~(\ref{J_dec}) and~(\ref{T_dec}) according to the Landau prescription and Landau matching, one has
\begin{equation}\label{dens_time_evol}
    \begin{split}
        0&=\partial_\mu J^\mu = \dot \rho + \rho \, \theta + \partial\cdot \nu, \qquad\qquad  \Rightarrow \qquad\qquad\dot \rho = -\theta \, \rho -\partial\cdot \nu,\\ \\
        0&=u_\nu \partial_\mu T^{\mu\nu} = \dot \ped + \left( \vphantom{\frac{}{}} \ped + \pres + \Pi \right) \theta - \pi : \sigma, \quad \Rightarrow\quad \dot \ped = -\theta\left( \vphantom{\frac{}{}} \ped + \pres + \Pi \right) + \pi : \sigma.
    \end{split}
\end{equation}
In the last equation, we made use of the Landau decomposition of the gradients of the four-velocity, namely
\begin{equation}
    \begin{split}
        \partial_\mu u_\nu &= u_\mu \, u^\alpha\partial_\alpha u_\nu + \nabla_\mu u_\nu = u_\mu \dot u_\nu +\omega_{\mu\nu} +\sigma_{\mu\nu} +\frac{1}{3} \,\theta \, \Delta_{\mu\nu}.
    \end{split}
\end{equation}
The acceleration term on the right-hand side is the only term proportional to the four-velocity. The purely orthogonal ones are divided into the antisymmetric $\omega_{\mu\nu}$ (relativistic vorticity), symmetric and traceless $\sigma_{\mu\nu}$ (shear flow), and traceful (and orthogonal to $u^\mu$) part proportional to the scalar expansion $\theta$, where
\begin{equation}
    \theta = \nabla\cdot u = \partial\cdot u.
\end{equation}
%
%\begin{equation}
%   \begin{split}
%        \theta &= \nabla\cdot u = \left( g^{\mu\nu} - u^\mu u^\nu \right) \partial_\mu u_\nu = \partial\cdot u - u^\nu(u\cdot\partial)u_\nu = \\
%        &= \partial\cdot u -\left[ (u\cdot \partial )(u\cdot u) - u_\nu(u\cdot\partial)u^\nu\right] =\partial\cdot u + u^\nu(u\cdot\partial)u_\nu = \partial\cdot u.
%    \end{split}
%\end{equation}
%
In obtaining the above identity, we have used the fact that the four-velocity is unitary ($u\cdot u =1$ is a constant, it differentiates to zero), hence $u^\nu(u\cdot\partial)u_\nu=-u^\nu(u\cdot\partial)u_\nu$ and it is therefore vanishing.

Making use of both Eqs.~(\ref{dot_T-mu}) and~(\ref{dens_time_evol}), one has
\begin{equation}
   \begin{split}
        \frac{\partial \pedeq}{\partial T} \, \dot T + \frac{\partial \pedeq}{\partial\mu} \dot \mu &= -\theta \left( \vphantom{\frac{}{}} \ped +\pres +\Pi  \right) +\pi : \sigma,\\
        \frac{\partial \rhoeq}{\partial T} \, \dot T + \frac{\partial \rhoeq }{\partial \mu} \, \dot \mu &= -\theta \, \rho -\partial\cdot \nu,
        \label{system}
    \end{split}
\end{equation}
Since the determinant of this linear system is the Jacobian in~(\ref{Jacobian}), which we assumed to be non-vanishing, the functions $\pedeq$, $\rhoeq$ and their partial derivatives are known and the effective temperature and chemical potential can be written in terms of the physical densities thanks to the Matching~(\ref{L_match}). It is possible to invert the system~(\ref{system}) and write a closed formula for the comoving derivatives in terms of the hydrodynamic degrees of freedom. It is a common practice in relativistic kinetic theory to make use of the variables
\begin{equation}\label{def_alpha_beta}
    \alpha  = \frac{\mu}{T}, \qquad \beta = \frac{1}{T}.
\end{equation}
Since there is a one-to-one relation between\footnote{The determinant of the Jacobian matrix of the transformation is $-\beta^3=-1/T^3\neq 0$, $\forall T>0$.} $\{\alpha, \beta\}\leftrightarrow \{ T,\mu \}$, it is possible to write the last system of equations in terms of the effective $\alpha$ and $\beta$, and their comoving derivatives
\begin{equation}\label{alph_beta_dot}
   \begin{split}
        \frac{\partial \pedeq}{\partial \alpha} \, \dot \alpha + \frac{\partial \pedeq }{\partial \beta} \, \dot \beta &=-\theta \left( \vphantom{\frac{}{}} \ped +\pres +\Pi  \right) +\pi: \sigma,\\
        \frac{\partial \rhoeq}{\partial \alpha} \, \dot \alpha + \frac{\partial \rhoeq }{\partial \beta} \, \dot \beta &= -\theta \, \rho -\partial\cdot \nu.
    \end{split}
\end{equation}
This system of equations can be solved similarly to the previous one, because its characteristic determinant reads
\begin{equation}
    \begin{split}
        0&\neq\frac{\partial \pedeq}{\partial T}\frac{\partial \rhoeq}{\partial \mu}- \frac{\partial \pedeq}{\partial \mu}\frac{\partial \rhoeq}{\partial T} = -\beta^2\left[ \left(\alpha\frac{\partial \pedeq}{\partial \alpha} + \beta \frac{\partial \pedeq}{\partial \beta}\right)\frac{\partial \rhoeq}{\partial \alpha}-\frac{\partial \pedeq}{\partial \alpha} \left(\alpha\frac{\partial \rhoeq}{\partial \alpha} + \beta \frac{\partial \rhoeq}{\partial \beta}\right) \right]
        \\
        &=\beta^3\left[ \frac{\partial \pedeq}{\partial \alpha}\frac{\partial \rhoeq}{\partial \beta} - \frac{\partial \pedeq}{\partial \beta}\frac{\partial \rhoeq}{\partial \alpha}\right], \quad \Rightarrow \quad \left[ \frac{\partial \pedeq}{\partial \alpha}\frac{\partial \rhoeq}{\partial \beta} - \frac{\partial \pedeq}{\partial \beta}\frac{\partial \rhoeq}{\partial \alpha}\right] = -T^3 \left[ \frac{\partial \pedeq}{\partial T}\frac{\partial \rhoeq}{\partial \mu}- \frac{\partial \pedeq}{\partial \mu}\frac{\partial \rhoeq}{\partial T} \right] \neq 0.
    \end{split}
\end{equation}
Introducing the notation
\begin{equation}\label{short-hand_partials}
    \ped_\alpha =\frac{\partial\pedeq}{\partial \alpha}, \qquad \ped_\beta =\frac{\partial\pedeq}{\partial \beta}, \qquad \rho_\alpha =\frac{\partial\rhoeq}{\partial \alpha}, \qquad \rho_\beta =\frac{\partial\rhoeq}{\partial \beta}, \qquad \qquad {\cal J} = \ped_\alpha\rho_\beta - \ped_\beta\rho_\alpha,
\end{equation}
the exact solutions of Eqs.~(\ref{alph_beta_dot}) are
\begin{equation}\label{alpha_beta_dot}
    \begin{split}
        \dot \alpha &= \left\{\left[ \theta \, \rho +\partial\cdot \nu \vphantom{\frac{}{}}\right] \ped_\beta -\left[ \theta\left( \vphantom{\frac{}{}}\ped +\pres + \Pi \right) -\pi:\sigma \right] \rho_\beta \right\}/{\cal J}, \\
        \dot \beta &= \left\{\left[ \theta\left( \vphantom{\frac{}{}}\ped +\pres + \Pi \right) - \pi:\sigma  \right] \rho_\alpha -\left[ \theta \, \rho +\partial\cdot \nu \vphantom{\frac{}{}}\right] \ped_\alpha\right\}/{\cal J}.
    \end{split}
\end{equation}
In a similar way, it is possible to write an exact relation between the spatial gradients of the effective parameters $\alpha$ and $\beta$ and the hydrodynamic degrees of freedom. 

Only the projection along the four-velocity of the four-momentum conservation was used to derive the exact relations~(\ref{alpha_beta_dot}). The orthogonal projection reads
\begin{equation}\label{pres_grad}
    \begin{split}
        0&=\Delta^\mu_\rho \partial_\sigma T^{\rho\sigma} = \left( \vphantom{\frac{}{}} \ped +\pres +\Pi  \right) \dot u^\mu -\nabla^\mu\left( \vphantom{\frac{}{}} \pres + \Pi  \right) + \Delta^\mu_\rho \partial_\sigma \pi^{\rho\sigma},\\
        &\Rightarrow \nabla^\mu \pres = \left( \vphantom{\frac{}{}} \ped +\pres +\Pi  \right) \dot u^\mu -\nabla^\mu\, \Pi + \Delta^\mu_\rho \partial_\sigma \pi^{\rho\sigma}.
    \end{split}
\end{equation}
Making use of the grand-canonical thermodynamic relations 
\begin{equation}\label{gran_can_rel}
    \frac{\partial \pres}{\partial \mu} = \rho, \qquad \frac{\partial \pres}{\partial T} = \frac{\ped +\pres -\mu \, \rho}{T}, \qquad \Rightarrow\qquad \left( T\frac{\partial}{\partial T} + \mu \frac{\partial}{\partial \mu} \right)\pres = \ped + \pres,
\end{equation}
it is possible to write the gradient of the hydrostatic pressure $\nabla^\mu\pres$ in terms of the gradients of the effective parameters $\alpha$  and $\beta$. Indeed,
\begin{equation}
    \begin{split}
        \nabla^\mu \pres &= \frac{\partial\pres}{\partial \alpha} \nabla^\mu \alpha + \frac{\partial\pres}{\partial \beta} \nabla^\mu \beta = \left[ \frac{\partial\pres}{\partial T} \frac{\partial T}{\partial \alpha} + \frac{\partial\pres}{\partial \mu} \frac{\partial \mu}{\partial \alpha} \right] \nabla^\mu \alpha + \left[ \frac{\partial\pres}{\partial T} \frac{\partial T}{\partial \beta} + \frac{\partial\pres}{\partial \mu} \frac{\partial \mu}{\partial \beta} \right] \nabla^\mu \beta \\
        &= \left[ \frac{1}{\beta} \, \rho \right] \nabla^\mu \alpha + \left[ \frac{\ped +\pres -\mu \, \rho}{T} \left( -\frac{1}{\beta^2} \right) + \rho \left( -\frac{\alpha}{\beta^2} \right)\right] \nabla^\mu \beta = -\frac{1}{\beta} \left( \vphantom{\frac{}{}} \ped +\pres \right) \nabla^\mu \beta + \frac{1}{\beta }\, \rho \, \nabla^\mu \alpha.
    \end{split}
\end{equation}
Plugging it into~(\ref{pres_grad}), one has
\begin{equation}\label{alpha-beta_grad}
    \nabla^\mu \beta = \left(\frac{\rho}{\ped+\pres}\right) \nabla^\mu \alpha -\beta \left(  1+\frac{\Pi}{\ped +\pres}\right) \dot u^\mu +\frac{\displaystyle \beta\left(\nabla^\mu \vphantom{\frac{}{}} \Pi  - \Delta^\mu_\rho \partial_\sigma \pi^{\sigma\rho}\right)}{\displaystyle \left( \vphantom{\frac{}{}}\ped + \pres  \right)}.
\end{equation}
The usefulness of the formulas~(\ref{alph_beta_dot}) and~(\ref{alpha-beta_grad}) becomes apparent when one has a deeper look at the usual way to extract hydrodynamics from a fundamental theory, and the needed phenomenological approximations and manipulations to obtain the actual functional relations between the transport coefficients.

The simplest assumption to solve for the hydrodynamics degrees of freedom is the relativistic perfect hydrodynamics which consists of the fluid flow velocity, the energy density, and the baryon number density. The other hydrodynamic degrees of freedom are assumed to be negligible. The five conservation equations~\eqref{conservation} are enough to solve the system for the three independent degrees of freedom for the four-velocity ($u\cdot u=1$ removes one), $\ped$, and $\rho$. The hydrostatic pressure is given by the equation of state, through the Landau matching~(\ref{L_match}) as a function of $\ped$ and $\rho$. The remaining $\pi^{\mu\nu}$, $\Pi$, and $\nu^\mu$ are assumed to be vanishing. Furthermore, one assumes that they are functionals of the other degrees of freedom, specifically gradients of them. The physical quantities in non-dissipative hydrodynamics are assumed to be the only ones of order zero in the gradient expansion~\footnote{This is not necessarily true for solids -- shear stress can be present at mechanical equilibrium in the absence of any gradient.}. Then, others are approximated at first order with the simple expressions (compatible with the geometric sector, symmetry, tracefulness, and orthogonality to $u^\mu$), for instance $\pi^{\mu\nu} = 2\eta(\ped,\rho) \sigma^{\mu\nu}$ or, equivalently $\pi^{\mu\nu} = 2\eta(T,\mu)\sigma^{\mu\nu}= 2\eta(\alpha,\beta)\sigma^{\mu\nu}$, and with more terms at higher orders. Besides the stability problems that this procedure entails, which are also present at the non-relativistic level, there is no clear way to extract the functional dependences of the transport coefficients (like $2\eta(T,\mu)$) without making further assumptions. Another popular method to extract hydrodynamics is to start from the grand-canonical equilibrium relation with the entropy density $s$
\begin{equation}\label{s^0_eq}
    s = \beta \left( \ped +\pres \right) - \alpha \rho,
\end{equation}
which corresponds to the entropy flux, 
\begin{equation}
    s^\mu = s u^\mu = \beta \left( \ped +\pres \right) u^\mu - \alpha \, \rho \, u^\mu = \beta \, \pres \, u^\mu +\beta \, T^{\mu\nu}u_\nu -\alpha J^\mu.
    \label{sflux}
\end{equation}
Namely, in the lab frame, the four-velocity is $u=(1,0,0,0)$ and the diffusion current vanishes because of the rotation invariance of the grand-canonical ensemble. Equation~\eqref{sflux} is therefore just Eq.~\eqref{s^0_eq} in a covariant notation. The basic assumption is that the entropy flux (\ref{sflux}) will read, at least for arbitrarily small deviations from the grand-canonical equilibrium
\begin{equation}
    s^\mu = \beta \, \pres(\alpha,\beta) \, u^\mu + \beta\,  T^{\mu\nu}u_\nu -\alpha J^\mu + \delta Q^\mu.
\end{equation}
The off-equilibrium current $\delta Q^\mu$ is at least second order in a combination of $\pi^{\mu\nu}$, $\Pi$, and $\nu^\mu$, and hence it does not change the structure of the first order contributions.
%bit it otherwise depends only on the hydrodynamic degrees of freedom. That is, 
Note, however, that the diffusion current from $J^\mu$ now is present anyway, because of the lack of rotation invariance for a generic (if small) deviation from equilibrium. Moreover, any function of $\alpha$ and $\beta$ in $\delta Q^\mu$ is allowed because $\alpha(\ped,\rho)$ and $\beta(\ped,\rho)$ do not depend on the dissipative currents owing to the Landau prescription. The non-negativity of the entropy production imposes additional bounds in the evolution of the system besides the ones given by the conservation equations~(\ref{conservation}). Indeed, making use of Eq.~\eqref{gran_can_rel}, we have
\begin{equation}
    0\le\partial \cdot s = \beta \, \pi:\sigma - \beta \, \theta \, \Pi - \nu \cdot \nabla  \alpha + \partial \cdot \delta Q.
\end{equation}
The simplest bound to ensure that the last inequality holds is that each term is separately non-negative. For arbitrarily small deviations from equilibrium, one can ignore the extra term in $\delta Q^\mu$, obtaining the first-order relations
\begin{equation}
    \pi^{\mu\nu} = 2\, \eta \, \sigma^{\mu\nu},\qquad \Pi = -\zeta \, \theta, \qquad \nu^\mu = \kappa_b \nabla^\mu \alpha,
\end{equation}
with all the transport coefficients positive. These relations are the same as the ones obtained from the point of view of the gradient expansion. On top of that one obtains the sign of them. Similarly, the second-order equations (keeping the $\delta Q$ at the lowest order, see Ref.~\cite{Muronga:2003ta}) result in the familiar relaxation-type equations. For instance for the bulk pressure $\Pi$ one has
\begin{equation}
   \tau_\Pi \dot \Pi + \Pi = -\zeta \, \theta - \frac{1}{2}\zeta \, T \left[\partial_\mu \left( \frac{\tau_\Pi}{2\zeta \, T} u^\mu \right)\right] \Pi + \lambda_{\Pi\nu} \nabla_\mu \nu^\mu.
\end{equation}
If $\tau_\Pi\to 0$ the second order equations get closer to the first order approximation $\Pi \to -\zeta \theta$, at least for small perturbations. A similar equation holds for the other terms. Nevertheless, even with this procedure the biggest practical problem with the gradient expansion approach is that it is exceedingly hard to extract all the second-order transport coefficients from a realistic interacting theory. For QCD it is already complicated to extract the equation of state in the presence of a non-vanishing baryon density (hence $\mu \neq 0$). It is therefore not practical at the moment to obtain all the transport coefficients from lattice calculations.

Relativistic kinetic theory is not really appropriate for a strongly interacting system, since it stems from quantum field theory in the classical (neglecting higher-order contributions in $\hbar$) and asymptotically weak interaction limits. However, it has the discrete advantage that the exact evolution of the hydrodynamic degrees of freedom has a lot of self-coupling from the very beginning, and the contribution of the non-hydrodynamic degrees of freedom can be treated systematically. This can be used to provide formulas for the second-order transport coefficients. On the other hand, it is incompatible though with a realistic equation of state. Besides the (relatively small) corrections due to the Bose or Fermi statistics of the different particle species, the equation of state is essentially ideal. All partial contributions to the pressure are essentially the temperature times the particle's density (exactly in the Boltzmann limit). The phenomenological ways to proceed imply breaking further the assumption of relativistic kinetic theory. For instance, obtaining the relations between the transport coefficients from an ideal gas. Writing them in terms of the familiar degrees of freedom of thermodynamics, like the equilibrium pressure, energy density, and the speed of sound, then using the relations between them from a realistic equation of state instead of the ideal gas one started with. Quasi-particles can be used to derive the relations but, as an additional complication, it is not possible to fix all the thermodynamic relations with medium-dependent (hence temperature, chemical potential) masses. It is necessary to add a medium-dependent bag term in the thermodynamics, and that in turn may require also out-of-equilibrium corrections, to ensure energy and momentum conservation. All of this will be discussed further in the remainder of this work. 

As it has been recently formalized, a lot of the desirable properties of relativistic kinetic theory in extracting hydrodynamics stem from the evolution equation of the Wigner distribution, the quantum precursor of the distribution function. It is not necessary to go to the full kinetic limit. Therefore in this work, we prefer to develop quasi-particle dynamics from a quantum field theory setup, rather than the historically more common phenomenological extension of the relativistic Boltzmann equation. This approach is also more flexible and easier to generalize, for instance, to include off-shell effects. Stemming from a general quantum field theoretical framework, one can simply relax some of the phenomenological assumptions rather than making new ones.

To do so, one must look back at the starting equations~(\ref{averages}), which must be plugged in~(\ref{conservation}). The operators corresponding to the stress-energy tensor $\hat T^{\mu \nu}$ and $\hat J^\mu$ can be derived from the lagrangian density of the system, which is always of the form
\begin{equation}
    \hat {\cal L}_0 +\hat {\cal L}_{\rm int}.
\end{equation}
The free part $\hat {\cal L}_0$ being the Klein-Gordon, Dirac densities, or their generalizations to higher spins. Following the procedure in Ref.~\cite{degroot}, both the scalar and spin $1/2$ contributions to the stress-energy tensor from the ${\cal L}_0$ can be written as an integral
\begin{equation}\label{T_0}
   T_0^{\mu\nu} = \int d^4 p \; p^\mu p^\nu W(x,p),
\end{equation}
which is almost identical to the formula in relativistic kinetic theory. The main difference being that $W$ is not on the mass shell, $p^2\neq 0$ in general, and it can be negative. The Wigner distribution $W$ is the quantum precursor of the distribution function. In the kinetic limit $W\to (2\pi)^{-3}\delta(p^2 -m^2) f$. In the simple case of a scalar field it reads
\begin{equation}\label{Wigner_tranform}
    W(x,p) = 2\int \frac{d^4v}{(2\pi)^4} \; e^{-i p\cdot v} \,{\rm tr}\left( \hat \rho  \; \hat \phi^\dagger(x+\tfrac{1}{2}v) \, \hat \phi(x-\tfrac{1}{2}v)\right).
\end{equation}
For fermions, it can still be expressed as the Fourier transform of the two-point function with respect to the state $\hat \rho$, but it also include a trace over the over the spinorial degrees of freedom. The two point function itself is a matrix. More often than not, the interaction part of the Lagrangian $\hat {\cal L}_{\rm int}$ has a non-vanishing contribution to the stress-energy tensor. Such a contribution does not depend on the scalar $W(x,p)$ presented before.

The conserved current (electric, baryon etc.) is rather simple for scalars since it  stems from ${\cal L}_0$ and does not have corrections due to the interactions
\begin{equation}\label{scalar_current}
    J^\mu = q\int d^4 p \; p^\mu W(x,p),
\end{equation}
as one could expect considering the kinetic limit. For higher spin the current cannot be exactly described by the same scalar appearing in the stress-energy tensor formula. For instance the scalar $W(x,p)$ stems from the scalar bilinear $\bar\psi \psi$ through the Wigner transform, not the vector $\bar \psi \gamma^\mu \psi$. The current of a spin $1/2$ field is built from the trace of the matrix version of the Wigner distribution with the gamma matrices, not the trace with the identity (scalar). The fermionic current can be approximated with~(\ref{scalar_current}) in some limits. More generally, relativistic kinetic theory can be recovered from quantum field theory making a semiclassical approximation for weakly interacting systems. This, in particular, entails also that the current of fermions reduces to the same formula with a scalar distribution function, just like in the scalar case.
 
We remind here that the main observations at the basis of the procedures in~\cite{degroot}, to obtain the relativistic kinetic theory, can be summed up to the evolution equation 
\begin{equation}\label{pre_kinetic}
    \frac{1}{4} \hbar^2 \partial\cdot \partial W-\left( \vphantom{\tfrac{1}{2}} p^2 -m^2 \right) W +i\hbar \, p\cdot \partial W = \cdots,
\end{equation}
The right-hand side depends on the interaction. This can be seen when applying the Klein-Gordon operator to the two-point expectation value 
\begin{equation}
    \left[ \vphantom{\tfrac{1}{2}}\partial_{(y)}\cdot\partial_{(y)} + m^2 \right]{\rm tr}\left( \hat \rho \; \hat\phi^\dagger(y)\, \hat\phi(z) \right) = \cdots,
\end{equation}
making the change of variables
\begin{equation}
    \begin{cases}
        y= x +\tfrac{1}{2}v, \\
        z = x - \tfrac{1}{2}v,
    \end{cases} \Rightarrow \begin{cases}
        \partial_\mu^{(y)}= \tfrac{1}{2}\partial_\mu^{(x)} +\partial_\mu^{(v)}, \\
        \partial_\mu^{(z)}= \tfrac{1}{2}\partial_\mu^{(x)} -\partial_\mu^{(v)}.
    \end{cases}
\end{equation}
and then considering the Fourier transform over the $v=y-z$ variables according to Eq.~(\ref{Wigner_tranform}). As mentioned before, this is valid not only for scalars, but all the Wigner transforms of two fields. That is all the real degrees of freedom for the matrix-valued, generalization of~(\ref{Wigner_tranform}) for higher spin fields. This is because, regardless of the spin, each component of a field fulfills a Klein-Gordon equation with sources
\begin{equation}
    \partial\cdot \partial \Psi_a + m^2 \Psi_a = \cdots.
\end{equation}
This can also be generalized to Wigner transforms besides the case of two fields of the same species, using almost the same procedures. Namely, starting from the expectation value of a generic combination of fields at the same point
\begin{equation}\begin{split}
    {\rm tr}\left( \hat \rho \; \hat\psi_a(x)\hat\psi_b(x)\cdots \right) &= \int d^4v \; \delta^4(v) \,{\rm tr}\left( \hat \rho \; \hat\psi_a(x +\tfrac{1}{2}v)\hat\psi_b(x-\tfrac{1}{2}v)\cdots \right)\\
    &=\int d^4v \; \int\frac{d^4 k}{(2\pi)^4} \; e^{-ik\cdot v} \,{\rm tr}\left( \hat \rho \; \hat\psi_a(x +\tfrac{1}{2}v)\hat\psi_b(x-\tfrac{1}{2}v)\cdots \right)
    \end{split}
\end{equation}
any kind of kinetic-like equation can be obtained for the field configurations, including the interaction terms. The problematic part is not finding equations that can be solved to obtain the evolution of the hydrodynamic quantities, but solving them. Or, more realistically, making some good approximation that can be computed without compromising the accuracy of the final results.

Relativistic kinetic theory can be obtained as a small $\hbar$ expansion. The zeroth order of Eq.~(\ref{pre_kinetic}) is just the on-shell constraint\footnote{Assuming that the interaction is at least linear in the Plank constant, as it can be the case because of dimensional reasons. However, in general, it must be checked on a case-by-case basis.} $W\propto \delta(p^2-m^2)$, then the first order is already in the form of relativistic kinetic theory. It is not the point of this work to discuss the derivation of the relativistic kinetic theory or its generalization to spinning particles. It suffice to note at this point, that the main idea of a semiclassical expansion is questionable for heavy-ion collisions. In the plasma before the hadronization, the action scale can be of the order of $\hbar \approx 200$ fm$\cdot$MeV$/$c or seven smaller. Average momenta of the order of hundreds of MeV$/$c, but rapid changes in small fractions of a fm. At least for a non-interacting scalar field, the quantum corrections have been recently computed and they are rather large~\cite{Tinti:2023mtv} compared to an ideal gas. There is no particular reason to assume an asymptotically weak interaction either: to explain the experimental data a strongly interacting liquid (closer to a perfect fluid) seems favored rather than weakly interacting one (closer to an ideal gas). An almost insurmountable obstacle to the adoption of kinetic theory in heavy ions is the selection of the degrees of freedom, the species of particles, and their masses. At the initial stage, at least for the perturbative (large momentum exchanged) regime, quarks and gluons seems the most reasonable degrees of freedom. However, hydrodynamics is used up to hadronization and thereafter mesons and baryons are the better approximation.

Hydrodynamics, on the other hand, does not try to reproduce accurately the full dynamics of the system. The objective is to reproduce the evolution of the (statistical average of) the locally conserved currents. The Copper-Frye prescription then translates them to particles. It can be used directly as a prediction, or additional scattering, and subsequent freeze-out may be considered. From this point of view, it makes sense to consider a single scalar that reproduces the physical current as in ~(\ref{scalar_current}). In the particular case of theory if interacting scalar and pseudoscalar fields no phenomenological approximations would be needed. The scalar is the sum of the Wigner distribution of each charged field, whatever they may look like. In the more general case one can always say that, regardless of the details of a theory, a scalar can be found that fulfills this equation. To be a usable approximation, some phenomenological simplifying assumptions must be added. In this work, we will consider only a scalar of the form
\begin{equation}
    W(x,p) = \delta(p^2 - M^2) W_{\rm on}(x,p),
\end{equation}
that is on-shell with respect to the medium dependent mass $M(\alpha,\beta)$. The effective $\alpha$ and $\beta$ are linked to the physical values of the Stress-energy tensor on the current vector as it was introduced before. Namely using~(\ref{def_alpha_beta}) and~(\ref{def_T_mu}), with the Landau definition of the four-velocity~(\ref{u_Landau}) and the decomposition~(\ref{J_dec}) of the current. We will address soon the definition of the in-medium effective mass $m$. At the moment, however, we remember that the on-shell condition is enough to split the on-shell part $W_{\rm on}$ in a positive and negative frequency. That is, without loss of generality and in a Lorentz covariant fashion
\begin{equation}
\begin{split}
    W(x,p) = \delta(p^2 - M^2) W_{\rm on}(x,p) = & 2\Theta(p_0) \, \delta(p^2-M^2)\, \frac{g_q}{(2\pi)^3}\, f^q(x,p) \\
    &+  2\Theta(-p_0)\, \delta(p^2-M^2) \, \frac{g_q}{(2\pi)^3} \, f^{\bar q}(x,-p),
\end{split}
\end{equation}
with $\Theta$ being the step function $\Theta(x>0)=1$, $\Theta(x<0)=0$. The factor of two and the $g_q/(2\pi)^3$ is there in analogy with the relativistic kinetic theory\footnote{$g_q/(2\pi \hbar)^3 = g_q/h^3$ if one does not employ the natural units and we use $g_q$ to denote degeneracy factor for particles as well as antiparticles.}. The awkward flip in the sign of the last term is also in analogy with kinetic theory. Indeed, for any Lebesgue integrable current
\begin{equation}\label{curent_antysymm}
    \begin{split}
        J^\mu &= q \int d^4 p \; p^\mu W(x,p) = \frac{1}{2} \, q \int d^4 p \; p^\mu W(x,p) + \frac{1}{2} \, q \int d^4 p \; p^\mu W(x,p) \\
        &= \frac{1}{2} \, q \int d^4 p \; p^\mu W(x,p) - \frac{1}{2} \, q \int d^4 p \; p^\mu W(x,-p) = \frac{1}{2} \, q \int d^4 p \; p^\mu \left[ W(x,p) - W(x,-p) \vphantom{\tfrac{1}{2}}\right].
    \end{split}
\end{equation}
Namely it is sensitive only to the antisymmetric part in the exchange $p\leftrightarrow -p$. Using the conventions introduced earlier, it reduces to the formula for gas of particles of charge $q$ (and antiparticles of charge $-q$)

\begin{equation}\label{J_q-qbar}
    J^\mu = q\int\frac{g_q \,d^3 p}{(2\pi)^3 E_{\bf p}(x)} \, p^\mu \left[ f^q(x,p) - f^{\bar q}(x,p) \vphantom{\tfrac{1}{2}}\right],
\end{equation}
where $E_{\bf p}(x)=\sqrt{M^2(\alpha,\beta)+{\bf p}\cdot{\bf p}}$ is the spacetime-dependent positive energy. The dependence on the spacetime point arises from $\alpha$ and $\beta$ in the effective mass. The notable difference with relativistic kinetic theory is that neither $f^q$ nor $f^{\bar q}$ have to be non-negative. They are a one-particle reduced probability density and have no particular probabilistic interpretation. In this work, for notational simplicity, we denote $f^{\pm}$ as the combination $f^q \pm f^{\bar q}$.

Up to now, the effective charged (anti) particle distribution functions have been introduced just as a pair of functions that are general enough to be able to correctly reproduce the physical baryon current $J^\mu$. Making them look like their relativistic kinetic theory counterparts is not just an aesthetic requirement. It is a convenient parametrization, that makes it possible to generalize the phenomenological assumption commonly used in relativistic kinetic theory. In particular, it makes it possible to extract the transport coefficients of second-order viscous hydrodynamics using the same mathematical tools used in kinetic theory. The first phenomenological assumption that we introduce, which does not constitute a convenient parametrization but an actual approximation, is to consider the kinetic-like equation
\begin{equation}\label{Bol_Vlas_-}
    p^\mu \partial_\mu f^- + \tfrac{1}{2}\partial_\mu M^2\,  \partial^\mu_{(p)}f^- = - \frac{(p\cdot u)}{\taueq}\left( \vphantom{\frac{1}{2}} f^- -\fnm \right).
\end{equation}
The right-hand side is clearly the collisional kernel treated in the relaxation-time approximation. The assumption here is that the combined effect of the interactions will send the system, as a first approximation, toward local equilibrium. We note that the local equilibrium here concerns the effective $f^-$, as opposed to the actual Wigner distributions of the physical degrees of freedom. The precise form of $\fnm$ will be discussed in the next section. At the moment, it suffices to say that, by definition, the current produced by $\fnm$ has no diffusion term $\nu^\mu$ regardless of the physical $J^\mu$, but coincides with the latter for the charge density. In other words
\begin{equation}\label{Jeq}
    J^\mu|_{\fnm}(x) = \rhoeq(\alpha,\beta) u^\mu.
\end{equation}
The physical interpretation is that the statistical forces mainly drive the system toward a local version of the global equilibrium relations.

The origin of the Boltzmann-Vlasov-like equation~(\ref{Bol_Vlas_-}) can be seen from the charge conservation equation. Expanding explicitly the derivatives and making use of the identity for the distributions
\begin{equation}\label{delta_prime_prop}
    2p^\mu \, \delta^\prime (p^2-M^2) = \partial^\mu_{(p)}\, \delta(p^2-M^2),
\end{equation}
one has
\begin{equation}\label{def_Vlas}
    \begin{split}
        0 = \partial_\mu J^\mu =& q\frac{g_q}{(2\pi)^3} \int d^4 p \; 2\Theta(p^0) \left[  -2 M \, p\cdot\partial M \, \delta^\prime(p^2-M^2) f^- + \delta(p^2 -M^2) \, p\cdot \partial f^- \vphantom{\frac{}{}}\right] \\
        &= q\frac{g_q}{(2\pi)^3} \int d^4 p \; 2\Theta(p^0) \, \delta(p^2-M^2) \, \left[ p\cdot \partial f^- + \tfrac{1}{2}\partial_\mu M^2 \, \partial^\mu_{(p)} f^- \right].
    \end{split}
\end{equation}
The conservation of the charge is ensured by the definition itself~(\ref{L_match}) of the effective temperature and chemical potential, hence $\alpha= \mu/T$ and $\beta= 1/T$
\begin{equation}
    \partial_\mu J^\mu = -q \frac{1}{\taueq} u_\mu \int \frac{g_q d^4 p}{(2\pi)^3 } \; 2 \Theta (p^0) \, \delta(p^2-M^2) \, \left[ \vphantom{\frac{}{}} f^- - \fnm \right] p^\mu = -\frac{1}{\taueq} \left[ \vphantom{\frac{}{}}  \rho - \rhoeq(\alpha, \beta)\right] = 0.
\end{equation}
Of course, the approximation is not complete yet. The definition of the effective $\alpha$ and $\beta$ passes through the physical energy density~(\ref{u_Landau}), hence the stress-energy tensor $T^{\mu\nu}$, which has not been addressed yet. The next phenomenological approximation is not necessary, strictly speaking, but it stems from physical considerations as well as convenience.

As it was seen earlier, in Eq.~(\ref{curent_antysymm}), the current is sensitive only to the antisymmetric part of a scalar weight $W(x,p)$ in the $p\leftrightarrow -p$ exchange. In particular, for an on-shell effective weight, on the subtraction $f^-$ of the (on-shell, positive energy) particle and antiparticle generalized distribution functions. The sum of the two, and more in general the symmetric part of a scalar weight $W$, is an independent degree of freedom. Without the classical restriction of a positive $f^q$ and $f^{\bar q}$, there is enough freedom to reproduce any value in space-time of an out-of-equilibrium $T^{\mu\nu}$ with a single, on-shell$f^+$, using the kinetic-like formula in~(\ref{T_0}). This is regardless of the actual value of the physical $T^{\mu\nu}$. Here we argue, however, that it might be more convenient to consider an extra scalar effective distribution. The motivation lies in the limit in which the relativistic Boltzmann equation is valid. That is a dilute, weakly interacting gas after hadronization. In this case there is a difference between a neutral fluid that consists of uncharged particles, pions, nearly massless, and another one made of an equal number of baryons and antibaryons, massive nucleons. A gas consisting mostly of pions will have a small trace anomaly $T^{\mu\nu} g_{\mu\nu}\approx m^2 \int W\to 0$, on the other hand, one consisting mostly of nucleons needs a significant deviation from local equilibrium to have a small trace anomaly. In this case, one would expect a fast evolution because of the relaxation time approximation. It is reasonable, then, to expect that a model with two generations of effective particles, one charged and one uncharged, with different effective masses, can more easily reproduce the evolution close to hadronization. For simplicity, we will consider the same relaxation time for both of the generations. Therefore, besides the equation~(\ref{Bol_Vlas_-}), we consider
\begin{equation}\label{Bol_vlas_0+}
    \begin{split}
        &p\cdot\partial f^+ +\tfrac{1}{2}\partial_\mu M^2\,  \partial^\mu_{(p)} f^+ = -\frac{(p\cdot u)}{\taueq}\left( \vphantom{\frac{}{}} f^+ - \fnp \right), \\
        & p\cdot\partial f +\tfrac{1}{2}\partial_\mu m^2 \, \partial^\mu_{(p)} f = -\frac{(p\cdot u)}{\taueq}\left( \vphantom{\frac{}{}} f - \fn \right).
    \end{split}
\end{equation}
The first equation can be combined with~(\ref{Bol_Vlas_-}) to give
\begin{equation}\label{Bol_vlas_qqbar}
    \begin{split}
        &p\cdot\partial f^{q} +\tfrac{1}{2}\partial_\mu M^2\,  \partial^\mu_{(p)} f^{q} = -\frac{(p\cdot u)}{\taueq}\left( \vphantom{\frac{}{}} f^{q} - \fnq \right), \\
        & p\cdot\partial f^{\bar q} +\tfrac{1}{2}\partial_\mu M^2\,  \partial^\mu_{(p)} f^{\bar q} = -\frac{(p\cdot u)}{\taueq}\left( \vphantom{\frac{}{}} f^{\bar q} - \fnqbar \right).
    \end{split}
\end{equation}
As mentioned earlier, only the symmetric part in $p\leftrightarrow -p$ contributes to a term like~(\ref{T_0})
\begin{equation}
    \begin{split}
        \int d^4 p \; p^\mu p^\nu \, W(x,p) &= \frac{1}{2}\int d^4 p \; p^\mu p^\nu \left[ \vphantom{\frac{}{}} W(x,p) + W(x,-p) \right],
    \end{split}
\end{equation}
and, following the same arguments used for the current, only the $f^+$ term of the weight on-shell with mass $m$ contributes; and, without loss of generality, the term in $f$ with mass $m_0$ can also be an integral over positive energies only. In the end, introducing the notation
\begin{equation}\label{def_int}
    \begin{split}
        \int_{\bf p} &= \frac{g}{(2\pi)^3}\int d^4 p \; 2\Theta(p_0) \, \delta(p^2-m^2),\\
        \int_{\bf q} &= \frac{g_q}{(2\pi)^3}\int d^4 p \; 2\Theta(p_0) \, \delta(p^2-M^2),
    \end{split}
\end{equation}
for the Lorentz covariant, on-shell, positive energy integrals of the uncharged (denoted with a ${\bf p}$) and charged (denoted with a $\bf q$) scalar effective degrees of freedom, respectively. The stress-energy tensor takes the form
\begin{equation}\label{T_quas}
    T^{\mu\nu}= \int_{\bf p} p^\mu p^\nu \, f \; + \; \int_{\bf q} p^\mu p^\nu \, f^+ \; +\; B^{\mu\nu}.
\end{equation}
The presence of a bag-like term $B^{\mu\nu}$ cannot be excluded. As it will be discussed further in the next section, it is not possible to exactly match the equation of state at global equilibrium in absence of $B^{\mu\nu}$. In other words, in presence of medium dependent masses $M(\alpha,\beta)$ and $m(\alpha, \beta)$, thermodynamic relations at global equilibrium are not satisfied only with equilibrium distributions $\feq^q$, $\feq^{\bar q}$ and $\feq$.
%
%Not with kinetic-inspired equilibrium $\feq^q$, $\feq^{\bar q}$ and $\feq$: that is, kinetic relations but with medium dependent masses $M(\alpha,\beta)$ and $m(\alpha, \beta)$. 
%
In this work and we consider the simplest solution to overcome this issue, as discussed in reference~\cite{Tinti:2016bav}. Here we use the fact that the independent degrees of freedom are only four and the local four-momentum conservation is enough to solve for them as an initial value problem. This allows us to extract the explicit coefficients of second-order hydrodynamics using the tools of relativistic kinetic theory.

As a matter of notation, the left-hand side of the Boltzmann-Vlasov equation is often written in a slightly different form than that in Eqs.~\eqref{Bol_vlas_qqbar}, i.e., only the derivatives in three-momentum are considered. This is in fact equivalent to the notation used in this work. In all of the generalized distribution functions, it is not really necessary to consider an explicit dependence in all of the four components of $p^\mu$, since only the on-shell restriction counts thanks to Eq.~(\ref{def_int}). Many authors prefer the convention of integrating first in the energy, to later consider only three-dimensional momentum integrals with the understanding that the $p^0$ appearing in the momentum integrals is the on-shell energy. Integrating out the energy in the Boltzmann-Vlasov equation results in a non covariant equation for the so called on-shell version of the distribution function, which depends only on threee momentum variables. For proof and additional information see Appendix~\ref{sec:app_on-shell}. In this work we prefer the manifestly covariant notation, since it is simpler to use for direct calculations and, in the end, the final expressions do not change.

Needless to say, one can obtain the same results using a more traditional approach. Starting from the classical theory, replacing the fixed masses of particles with medium-dependent ones. After realizing that more degrees of freedom are needed, introducing the bag-like term to respect the thermodynamic relations. Further, introducing an off-equilibrium correction to the bag term to ensure that, in general, the four-momentum and (as well as baryon number) currents are locally conserved.

We opted to have a longer and more general presentation of the model for two main reasons. The first point is that all of the approaches necessary to generalize relativistic kinetic theory for quasi-particles with a bag tensor constitute a further breaking of the assumptions necessary for kinetic theory itself. This is on top of the already broken assumption of asymptotically weak coupling. Besides the smaller number of approximations, essentially the relaxation-type evolution with a single relaxation time, another advantage is to have a clearer picture of how to improve the treatment, if required. Not only replacing the relaxation-type evolution with a different generalized collisional kernel (as it can be done in kinetic theory), but also quantum field-induced changes. For instance, reorganizing the bag term if an effective division between kinetic-like and bag-like term (eg. semiclassical coherent fields) can be highlighted from a more microscopic model. Up to the use of a full off-shell treatment, which might very well include the additional structure from the spin. We will leave, however, all of these generalizations to future research.

%*************************** 
\section{Quasiparticle thermodynamics}\label{sec:therm}
%***************************
%
We introduced in the last section the basics of the framework required to extract second-order viscous hydrodynamics and its transport coefficients. The method is essentially an extension of the work~\cite{Tinti:2016bav}. The main difference is the presence of two quasiparticle species, charge carriers and neutral. This section and the next ones are dedicated to the generalization of the method due to the extra degrees of freedom and the extra conserved current considered, namely, the local conservation of baryons. The relaxation type equations~(\ref{Bol_Vlas_-}) and (\ref{Bol_vlas_0+}) depend on the equilibrium version of the generalized distribution functions. The four-velocity $u^\mu(x)$, the effective temperature $T(x)$ and chemical potential $\mu(x)$ for a generic configuration are defined through~(\ref{u_Landau}) and (\ref{eff_T-mu}). This section is dedicated to the global (grand-canonical) equilibrium properties that are used to define the effective $T$ and $\mu$, or, equivalently the $\alpha=\mu/T$ and $\beta=1/T$, that appear in the local equilibrium $\feq$, $\feq^\pm$ and contribute the approximated dynamics through the collisional relaxation type kernel. The grand canonical equilibrium can be summed up by the following formula for the density matrix
\begin{equation}
    \hat \rho = \hat \rho_{\sf GCE} = \frac{e^{-\frac{1}{T}\hat H+ \frac{\mu}{T} \hat Q}}{Z}, \qquad Z = {\rm tr}\left( e^{-\frac{1}{T}\hat H+ \frac{\mu}{T} \hat Q} \right).
\end{equation}
In this work, we will not discuss the details of how to properly renormalize interacting theories for this formula, essentially a thermodynamic infinite-volume limit, to make sense of the extraction of equation of state. The important thing is that the renormalized Hamiltonian $\hat H$ and baryon number operator $\hat Q$ are supposed to commute between themselves and the linear and angular momentum operators $\hat {\bf P}$, $\hat {\bf J}$, as it is expected in a Lorentz covariant theory. In other words, the system is invariant by space-time translation and space rotations, as the formula suggests.

Because of this, the global equilibrium expectation value of the hydrodynamic tensors~(\ref{averages}) are characterized by only three independent scalars. Therefore, defining $t^\mu=(1,0,0,0)$ as the time direction, one has
\begin{equation}
    T^{\mu\nu}_{\sf GCE}= \ped_{\sf GCE}(T,\mu) \; t^\mu t^\nu -\pres_{\sf GCE}(T,\mu)\left( g^{\mu\nu} - t^\mu t^\nu \right), \qquad J^\mu_{\sf GCE}= \rho_{\sf GCE}(T,\mu) \; t^\mu.
\end{equation}
For the rest of this work, we will consider exclusively the grand-canonical global equilibrium, and we will use the lower $eq$ to indicate both the local and global equilibrium. It should be clear by the context when it is used for the static considerations of thermodynamics or the dynamics for its local version. It is important to note that the last decomposition holds for a generic homogeneous state (translation and rotation invariant) that can be chosen as an asymptotic state that the system approaches; in general\footnote{Even if we don't have any reason to suggest alternative homogeneous states, they are not hard to imagine, for instance, substituting the exponential with $e^{-\left( \hat H/T -\mu \hat Q /T  \right)^2}$, or any other bounded expression. The two parameters can still be used to fit the energy density and charge density. The pressure would not be, in general, equal to the global equilibrium state, and the higher moments of the observables would be different too.} there might be more degrees of freedom and intensive parameters. For instance, even in global equilibrium when the electric charge and isospin chemical potentials are added, other conserved currents should be considered in such cases. We will not consider this situation to avoid cumbersome algebra, however it is straightforward to generalize the presented framework.
%additional, unnecessary, complications to mathematical passages, but the concept are essentially the same.

Employing the subscript $0$ to denote the quasiparticle expectation values with respect to $\fn$ and $\fnpm$, from~(\ref{J_q-qbar}) and~(\ref{T_quas}) one has the equilibrium relations
\begin{equation}
    \Jeq^\mu =J_0^\mu, \qquad \Teq^{\mu\nu} = \Tk^{\mu\nu} +  B^{\mu\nu}_{\rm eq}.  
\end{equation}
The first one is rather immediate
\begin{equation}
    J_0^\mu= \int_{\bf q} p^\mu \fnm = \Jeq^\mu = \rhoeq(\alpha,\beta) \, t^\mu, \quad \Rightarrow \quad \rho_0 =\rhoeq,
\end{equation}
the equilibrium $\fnm$ must fit the global equilibrium density by itself.

Regarding the stress-energy tensor, the situation is more complicated. Some choices must be made about $B^{\mu\nu}_{\rm eq}$ as well as $\fnp$ and $\fn$. In this work, we will do as in Ref~\cite{Tinti:2016bav} and consider a diagonal equilibrium bag tensor
\begin{equation}
    \Tk^{\mu\nu} = \int_{\boldsymbol{p}} \fn\, p^\mu p^\nu + \int_{\boldsymbol{q}} \fnp\, p^\mu p^\nu~,~~~B^{\mu\nu}_{\rm eq} = B_0(\alpha\beta)\, g^{\mu\nu}.\label{equilibriummeanfield}
\end{equation}
There is one extra degree of freedom to fit compared to the case without a chemical potential~\cite{Tinti:2016bav}, namely the baryon density, but there is also one extra function, two masses $m$ and $M$. Regarding the specific forms of the equilibrium $\fn$, $\fnpm$, since the results obtained from kinetic theory are rather good, we argue that a kinetic-theory-inspired guess is going to be adequate also if one needs to consider the baryon current. Given the separation between charged and uncharged quasiparticles it might seem reasonable to use a Bosonic distribution for the $\fn$ and a Fermionic one for $\fnq$ and $\fnqbar$, in analogy with quarks and gluons. In this work, however, we prefer to use the Jutner forms
\begin{equation} 
         \fn= e^{-\beta(p\cdot u)}, \quad    \fnq= e^{q \alpha}e^{-\beta(p\cdot u)}, \quad   \fnqbar= e^{-q \alpha}e^{-\beta(p\cdot u)}.
       \nn\\  \label{dist}
\end{equation}
%
%If the mesons, as well as the gluons, are always Bosonic\footnote{But not always scalar or pseudoscalars, anyway.}, the baryons are not necessary fermions. 
More importantly, besides the extreme limits of asymptotic freedom and hadron resonance gas, it is hard to say what would be the more appropriate degrees of freedom and how many of them there are. There is no particular reason to believe that changing to Bosonic-Fermionic equilibrium distributions would improve the approximation better than changing the effective number of degrees of freedom $g$ and $g_q$ in the measure~(\ref{def_int}) or the charge $q$. In any case, the modifications to the final formulas for the Bose-Fermi statistics are rather immediate, compared to other, deeper, modifications. We will just keep the Jutner form, for simplicity and ease of reading of the formulas. From Eq.~(\ref{equilibriummeanfield}), total energy density and pressure in equilibrium are defined as 
\begin{equation} 
{\cal E}_{{\rm eq}} = u_\mu T_{{\rm eq}}^{\mu\nu} u_\nu, \quad {\cal P}_{{\rm eq}} = -\frac{1}{3}\Delta_{\mu\nu} T_{{\rm eq}}^{\mu\nu} ,
\end{equation}
with $u^\mu =t^\mu$ at global, rather than local, equilibrium. Hence, using Eqs.~(\ref{dist}), we may write
\begin{equation}
    \Teq^{\mu\nu} = \left( \ped_0 +B_{0} \right) u^\mu u^\nu -\left( \pres_0 -B_{0} \right) \Delta^{\mu\nu} \equiv \pedeq u^\mu u^\nu -\preseq \Delta^{\mu\nu},  \nn\\  
\end{equation}
where the kinetic part of energy and pressure  are read in terms of the modified Bessel functions $K_n(x)$~\cite{gradshteyn2007}
        \begin{align}
            \ped_0(T,\mu) &= \frac{g_q}{\pi^2}\cosh(q\,\alpha) T^4 \left[ \vphantom{\frac{}{}} 3 z_q^2 K_2(z_q) +z_q^3 K_1(z_q) \right] \nonumber\\&+ \frac{g}{2\pi^2} T^4 \left[ \vphantom{\frac{}{}} 3 z^2 K_2(z) +z^3 K_1(z) \right] , \\
            \pres_0(T,\mu) &= \frac{g_q}{\pi^2}\cosh(q\, \alpha) T^4 \left[ \vphantom{\frac{}{}}  z_q^2 K_2(z_q) \right]\nonumber\\&+ \frac{g}{2\pi^2}T^4 \left[ \vphantom{\frac{}{}}  z^2 K_2(z) \right], 
        \end{align}
with $z_q=\beta M(\alpha\beta)=M(\mu,T)/T$ and $z= \beta m(\alpha,\beta)= m(\mu,T)/T$. In a similar way, the baryon density reads
\begin{equation}
    \rhoeq = u_\mu \Jeq^\mu= \rho_{0}(T,\mu)=\frac{q g_q}{\pi^2}\, \sinh( q\, \alpha)\, T^3 \, z_q^2 K_2(z_q).\nn\\
\end{equation}
Since the function $x^2K_2(x)\in(0,2),\, \forall x>0$ is monotonically decreasing, it is always invertible. As long as $\rhoeq(\alpha\beta)$ does not go to zero slower than linearly in $\alpha$ (hence $\rhoeq/\sinh(\alpha)$ is limited, as it is usually the case\footnote{It is hard to say anything about the exact equation of state of QCD, so there could be a singular point in $\alpha=0$, but usually the realistic equation of state are given in a Taylor expansion around a vanishing chemical potential, $\rhoeq/\sinh(\alpha)$ has no pole in $0$.}) and the product $qg_q$ is large enough that the right-hand side is always smaller than $2$, the equation
\begin{equation}
   z_q^2 K_2(z_q) = \frac{\pi^2\beta^3 }{qg_q} \frac{\rhoeq(q\, \alpha,\beta)}{\sinh(\alpha)},
\end{equation}
can be solved for $z_q = \beta M(\alpha\beta)$, hence the effective mass $M$ is well defined. At this point we follow what was done previously in Ref~\cite{Tinti:2016bav} and define the mass $m(\alpha,\beta)$ fitting $\pedeq +\preseq = \ped_0 +\pres_0$, which does not depend on the bag $B_0$, then $B_0$ can be obtained from $\pedeq -\preseq= 2B_0 +$, a now known function in $\alpha,\beta$. The equilibrium bag term $B_0$ can be positive or negative and does not constitute a problem. To show that $m(\alpha,\beta)$ can be well defined, we need to use the fact that fitting $\rhoeq $ with $\rho_0$ constrains the product $qg_q$, but not $q$ itself. To be more specific, making simple algebraic manipulations to $\ped_0 + \ped_0 =\pedeq +\preseq$, similar to the ones used for $\rho_0 =\rhoeq$, one has
\begin{equation}\label{m_fix}
   4z^2 K_2(z) + z^3K_1(z) = \frac{2\pi^2 }{g} \left[ \beta^4\left(\pedeq(\alpha,\beta) +\preseq(\alpha,\beta)\vphantom{\frac{}{}}\right) - \frac{g_q}{\pi^2 }\cosh(q\,\alpha)\left( 4z_q^2 K_2(z_q) + z_q^3K_1(z_q)\vphantom{\frac{}{}}\right)\right],
\end{equation}
The left-hand side is monotonically decreasing from $8$ to $0$ for positive $z$. It makes sense to invert the formula, hence solving for $z=\beta\,  m(\alpha,\beta)\to m(\alpha,\beta)$, if the term on the right-hand side is also positive and within the interval $(8,0)$. The overall factor $(2\pi^2)/g$, hence $g>0$, can be fixed to take care of the magnitude, but it can't deal with the positivity. In the square bracket, there is a subtraction and the second term is defined as negative, for the very same mathematical reason the term on the left-hand side is positive. Having fixed $q g_q$ to ensure the well definition of the effective mass $M(\alpha,\beta)$, does not fix $g_q$ itself. It can be chosen to be arbitrarily small, fixing then $q$ to a larger number, without changing $M$ and then $z-q=\beta M$. The negative term in the square brackets of Eq.~(\ref{m_fix}) can be thus made arbitrarily small without loss of generality. The only type of equations of state that cannot be fitted by this procedure is an awkward dark-energy-like sector~\cite{Jaiswal:2023tkh}, in which, at least in some regions, the pressure is negative and equal in magnitude to the energy density or something even more extreme in which $\pedeq+\preseq$ can be negative. Needless to say, if one needs to fit these kind of equations of state, the method proposed in this work is inappropriate. However, for the regular candidates for a realistic equation of state, in which $\pedeq+\preseq$ is strictly positive, the $\alpha,\beta$-dependent $M$, $m$, and $B_0$ are well defined, given a judicious choice of the constants $g$, $g_q$ and $q$.

From the practical point of view, it can be envisioned that the larger the region of $\alpha$ and $\beta$ the more such constraints on the constants can be important. On the other hand, if the region is more limited it can be possible to choose them in such a way as to keep $m$, $M$, and $B_0$ within some boundaries, which might be useful for numerical considerations (eg. having an easier time making numerical estimates), or to follow some theoretical or phenomenological expectations.

We finish this section with a formula that will be used in the next section to simplify the dynamics, before any further approximation. it stems from the Gibbs-Duhem equation. Namely, the differential of the equilibrium pressure $\preseq(T,\mu)$ as a function of the intensive parameters
\begin{equation}\label{Gibbs_Duhem}
    d\preseq(T,\mu) =\beta\left( \pedeq(T,\mu) + \preseq(T,\mu) - \alpha \,\rhoeq(T,\mu) \vphantom{\frac{}{}} \right)dT \;+\; \rhoeq(T,\mu) \, d\mu, 
\end{equation}
or, equivalently, for the partial derivatives of $\pedeq$
\begin{equation}\label{Gibbs_Duhem_partial}
    \begin{split}
        \frac{\partial\preseq}{\partial T} &=\frac{ \pedeq(T,\mu) + \preseq(T,\mu) - \mu \,\rhoeq(T,\mu)}{T} \vphantom{\frac{}{}}, \\
        \frac{\partial\preseq}{\partial \mu}&=\ \rhoeq(T,\mu).
    \end{split}
\end{equation}
The interesting part is that the same relations would hold between $\pres_0$, $\ped_0$ and $\rho_0$ if the masses were constant since they are the formulas for a perfect gas of charged particles, their antiparticles, and uncharged particles. The differential $d\pres_0(T,\mu)$ takes an extra term in the partial derivatives of the masses, namely
\begin{equation}\label{dP_0}
    \begin{split}
        d\pres_0(T,\mu) &=\beta\left( \ped_0(T,\mu) + \pres_0(T,\mu) - \alpha \,\rho_0(T,\mu) \vphantom{\frac{}{}} \right)dT \;+\; \rho_0(T,\mu) \, d\mu \;+\:  \frac{\pres_0}{\partial m } \, dm \;+\: \frac{\pres_0}{\partial M }\, dM  \\
        &= d\preseq  \;+\:  \frac{\pres_0}{\partial m } \, dm \;+\: \frac{\pres_0}{\partial M }\, dM.
    \end{split} 
\end{equation}
The last equality follow from the fact that $\ped-0+\pres_0 =\pedeq+\preseq$ and $\rho_0=\rhoeq$ by construction, regardless of the thermodynamic parameters. Since $\preseq = \pres_0 -B_0$, one has
\begin{equation}
    d\preseq = d\pres_0 - dB_0 = d\preseq  \;+\:  \frac{\pres_0}{\partial m } \, dm \;+\: \frac{\pres_0}{\partial M }\, dM \:-\: dB_0,
\end{equation}
and therefore
\begin{equation}\label{dB_0}
    \frac{\pres_0}{\partial m } \, dm \;+\: \frac{\pres_0}{\partial M }\, dM \:-\: dB_0 =0.
\end{equation}
It is convenient to write the partial derivatives with respect to the masses direct integrals of $\fn$ and $\fnp$. We will show only the derivative with respect to $m$ since the other term is essentially the same
\begin{equation}
    \begin{split}
        \frac{\partial \pres_0}{\partial m} &= -\frac{1}{3}\Delta_{\mu\nu}\, \frac{\partial}{\partial m}\left[\frac{g}{(2\pi)^3}\int  d^4p \; 2\Theta(p_0) \delta(p^2-m^2) \, p^\mu p^\nu \, \fn\right] \\
        &= \frac{g}{(2\pi)^3}\left( -\frac{1}{3}\Delta_{\mu\nu} \right) \int  d^4p \; 2\Theta(p_0) \left[  -m\, 2p^\mu\delta^\prime(p^2-m^2) \vphantom{\frac{}{}} \right]\, p^\nu \, \fn\\
        &=m\left( -\frac{1}{3}\Delta_{\mu\nu} \right) \int_{p} \left[ \frac{\partial}{\partial p_\mu} \left( \vphantom{\frac{}{}} p^\nu \, \fn \right) \right] \\
         &=m\left( -\frac{1}{3}\Delta_{\mu\nu} \right) \left[ \int_{p} \left( g^{\mu\nu} \, \fn -p^\nu \beta^\mu \, \fn \vphantom{\frac{}{}}\right) \right]\\
         &= -m \int_{\bf p} \fn.
    \end{split}
\end{equation}
Equation~(\ref{dB_0}) then reads
\begin{equation}
    dB_0 \:+\: m\,dm \int_{\bf p}\fn  \: +\; M\, dM \int_{\bf q} \fnp =0,
\end{equation}
or equivalently, for the partial derivatives
\begin{equation}\label{dB0_partial}
    \begin{split}
        & \frac{\partial B_0}{\partial T} \:+\: m\frac{\partial m}{\partial T} \int_{\bf p}\fn  \: +\; M\frac{\partial M}{\partial T} \int_{\bf q} \fnp =0, \\
        & \frac{\partial B_0}{\partial \mu} \:+\: m\frac{\partial m}{\partial \mu} \int_{\bf p}\fn  \: +\; M\frac{\partial M}{\partial \mu} \int_{\bf q} \fnp =0.
    \end{split}
\end{equation}
The above relations are valid in local equilibrium as well as in global equilibrium. Multiplying the first of the~(\ref{dB0_partial}) by $\partial^\nu T$ and the second by $\partial^\nu \mu$ one has
\begin{equation}\label{thermocondition}
    \partial^\nu B_0 \:+\: m\, \partial^\nu m \int_{\bf p}\fn  \: +\; M\, \partial^\nu M \int_{\bf q} \fnp =0,
\end{equation}
The above formula is a nontrivial consequence of the thermodynamic relations of the grand-canonical ensemble and the definition of $m$, $M$, and $B_0$. Since the choice of the grand-canonical ensemble as the equilibrium distribution that the fluid cells tend to asymptotically is not mandatory. And since the last relation is going to be used to simplify a term on the dynamics that depends only on the effective $\alpha$ and $\beta$ and their derivatives, hence $T$ and $\mu$, hence the physical $\ped$ and $\rho$. It is important to note that choosing a different form of global equilibrium requires a reformulation of the method starting from this section.

At the moment, it is hard to get realistic equations of state. The use of homogeneous equilibrium ensembles\footnote{Translation and rotation invariants, eg. the micro-canonical, canonical, and grand-canonical equilibrium.} is dictated primarily by practical considerations. Only recently there have been some exact results for free fields in non-homogeneous equilibrium (eg. rotations). Differently from perfect gasses, quantum fluids do not have the diagonal form of the stress-energy tensor. Therefore $T^{\mu\nu}$ is not restricted to be proportional only to $u^\mu u^\nu$ and $g^{\mu\nu}$, as it happens for a classical gas in non-homogenous equilibrium. Also, it is not an effect related to spin, as free scalar fields are not an exception~\cite{Becattini:2020qol}. At the moment it is not clear if these effects from non-homogeneous equilibrium should be included in the formulation of local equilibrium or not. We will consider in this work only the simplest case, that the homogeneous equilibrium is a good enough approximation. However, we acknowledge that it might not be the most general situation. 

\section{Nonequilibrium quasiparticle dynamics}\label{sec:dynamics}
%***************************
%
This section is the last missing piece of the approximation. Namely, the dynamical equations for the bag term $B^{\mu\nu}$. The treatment is the same as in Ref.~\cite{Tinti:2016bav}. The difference is the two kinetic parts in Eq.~(\ref{T_quas}), instead of just one in the previous work
\begin{equation}
\begin{split}
    T^{\mu\nu} &= \int_{\bf p} f \; p^\mu p^\nu + B^{\mu\nu}, \quad \longrightarrow \quad T^{\mu\nu} = \int_{\bf p} f \; p^\mu p^\nu + \int_\bq f^+ p^\mu p^\nu + B^{\mu\nu}, \label{EMT}
\end{split}
\end{equation}
the conserved baryon current~(\ref{J_q-qbar}) and, of course, the equation of state with a non-vanishing baryonic chemical potential discussed in the previous section. Many of the considerations remain unchanged. For ease of reading, we will repeat some of the arguments rather than referring to the literature and listing the (few) updates necessary for the generalized environment.

Different from the baryon number conservation, the Landau conditions ~(\ref{u_Landau}), and the kinetic equations~(\ref{Bol_Vlas_-}),~(\ref{Bol_vlas_0+}) do not ensure the energy and momentum conservation. To see that one can start expanding the four-momentum conservation equations, making use of the distribution identity~(\ref{delta_prime_prop})
\begin{equation}\label{cons_B}
    \begin{split}
        0&=\partial_\mu T^{\mu\nu} = \partial_\mu B^{\mu\nu} + \int_{\bf p} p^\nu \left[ p\cdot f + m\, \partial_\mu m \, \partial^\mu_{(p)} f  \right] + m\, \partial^\nu m\, \int_{\bf p}f \; +\int_{\bf q} p^\nu \left[ p\cdot f^+ + M\, \partial_\mu M \, \partial^\mu_{(p)} f^+  \right] + M\, \partial^\nu M\, \int_{\bf q}f^+ \\
        &= \partial_\mu B^{\mu\nu}  -\frac{u_\mu}{\taueq}\left[ \left( T^{\mu\nu} -B^{\mu\nu} \vphantom{\frac{}{}} \right) - \left( T_{\rm eq}^{\mu\nu} -B_{\rm eq}^{\mu\nu} \vphantom{\frac{}{}} \right)  \right] + m\, \partial^\nu m\, \int_{\bf p}f \; + M\, \partial^\nu M\, \int_{\bf q}f^+ \\
        &= \partial_\mu B^{\mu\nu} -\frac{u_\mu}{\taueq}\delta B^{\mu\nu}+ m\, \partial^\nu m\, \int_{\bf p}f \; + M\, \partial^\nu M\, \int_{\bf q}f^+.
    \end{split}
\end{equation}
Here, the $\delta B^{\mu\nu}$ is clearly the subtraction of the bag $B^{\mu\nu}$ with its local equilibrium counterpart $B^{\mu\nu}_{\rm eq} = B_0 g^\mu\nu$. The right-hand side can be further simplified using the relation~(\ref{thermocondition}), obtaining an equation involving only $\delta B^{\mu\nu}$ and the nonequilibrium parts of the generalized distributions $\delta f +f -\fn$, $\delta f^+= f^+-\fnp$ 
\begin{equation}
    \begin{split}\label{cons_deltaB}
        \partial_\mu \delta B^{\mu\nu} +\frac{u_\mu}{\taueq}\delta B^{\mu\nu}+ m\, \partial^\nu m\, \int_{\bf p}\delta f \; + M\, \partial^\nu M\, \int_{\bf q}\delta f^+ =0.
    \end{split}
\end{equation}
The last equation cannot be automatically satisfied if $\delta B^{\mu\nu} =0$. The effective masses depend exclusively on $\alpha =\mu/T$ and $\beta=1/T$, hence on the energy and baryon number density alone. Therefore, both $m\, \partial^\nu m$ and $M\, \partial^\nu M$ are fully described by $\ped$ and $\rho$ and their derivatives. They cannot be set to zero. On the other hand, as long as the bag $B^{\mu\nu} = B_0 (\alpha,\beta)g^{\mu\nu}$ is supposed to also depend exclusively on them, the only way to preserve the energy and momentum conservation is to assume that the integrals of $\delta f$,~$\delta f^+$ are vanishing. Besides, the complications of enforcing such a constraint at the dynamical level and how to modify the hydrodynamic expansion to take it into account, these conditions can be excluded on a physical basis. Namely, it would make the bulk pressure correction $\Pi$ vanishing
\begin{equation}
    \begin{split}
        \Pi &= -\frac{1}{3}\Delta_{\mu\nu}T^{\mu\nu}- \preseq = -\frac{1}{3}\Delta_{\mu\nu}\left(\int_{\bf p}p^\mu p^\nu f + \int_{\bf q}p^\mu p^\nu f^+ \right) -B_0 -\left( \pres_0 -B_0 \right) \\
        &=-\frac{1}{3}\left( g_{\mu\nu} - u_\mu u_\nu \vphantom{\frac{}{}}\right)\left[\int_{\bf p}p^\mu p^\nu \delta f + \int_{\bf q}p^\mu p^\nu \delta f^+ \right] = -\frac{1}{3}\left\{\int_{\bf p}\left[\vphantom{\frac{}{}} m^2 - (p\cdot u)^2\right] \delta f + \int_{\bf q}\left[\vphantom{\frac{}{}} M^2 - (p\cdot u)^2\right] \delta f^+ \right\}\\
        &= -\frac{1}{3}\left[  m^2 \int_{\bf p}\delta f + M^2 \int_{\bf q}\delta f^+ -\left( \ped - B_0 \vphantom{\frac{}{}} \right) +\left( \ped_0 \vphantom{\frac{}{}} \right)\right]\\
        &= -\frac{1}{3}\left(  m^2 \int_{\bf p}\delta f + M^2 \int_{\bf q}\delta f^+ \right).
    \end{split}
\end{equation}
In this work, following Ref.~\cite{Tinti:2016bav}, we will enjoy the freedom to choose the form of $B^{\mu\nu}$ out of equilibrium. Any rank two tensor, including the physical $T^{\mu\nu}-B^{\mu\nu}$ (with a generic $B^{\mu\nu}$) can be described by the kinetic part, reminding that, differently for their local equilibrium counterparts $\fn$ and $\fnp$, the $f$, and $f^+$ don not have to be positive\footnote{We remind here that the quasiparticles are not stemming from a classical theory that approximates the physical system. They represent a weight for the momenta that can reproduce the physical tensors of interest. The approximation stands in the dynamical equations that may not, in principle, reproduce adequately the evolution of the system.}. Since there are four independent equations in Ref.~(\ref{cons_deltaB}), we choose four additional\footnote{On top of the single scalar $B_0$, discussed in the previous section.} degrees of freedom for the general $B$ tensor 
\begin{equation}\label{Bnoneq}
B^{\mu\nu} = B^{\mu\nu}_{\rm eq} + \deltaB^{\mu\nu}~,~~~ \deltaB^{\mu\nu} = b_0 \,  g^{\mu\nu} + u^\mu b^\nu + b^\mu u ^\nu,
\end{equation}
so that the condition~(\ref{cons_deltaB}) necessary to ensure the local conservation of four-momentum is enough to solve for them as an initial value problem. The degrees of freedom are four because the vector $b^\mu$ is orthogonal to the fluid four-velocity, \textit{i.e.}~\mbox{$b\cdot u=0$}.

Plugging in the non-equilibrium bag part (\ref{Bnoneq}) in~(\ref{cons_deltaB}) and taking the projections proportional and orthogonal to the four-velocity one has the relaxation-type equations
\begin{equation}\label{deltaB_evol}
    \begin{split}
        \dot b_0 +\frac{1}{\taueq}b_0 \;=& \, b\cdot \dot u -(\partial\cdot b) - M\,\dot M \int_{\bf q} \delta f^+ -m\,\dot m \int_{\bf p}\delta f,\\ 
         \dot b^{\langle\mu\rangle} +\frac{1}{\taueq}b^\mu \; =& -\nabla^\mu b_0 -\theta b^\mu -b\cdot\partial \, u^\mu - M\nabla^\mu M \int_{\bf q}\delta f^+ -m\nabla^\mu m  \int_{\bf p}\delta f,
    \end{split}
\end{equation}
the immediate generalization of the equations found in Ref.~\cite{Tinti:2016bav}. In this work, we will consider second-order viscous hydrodynamics as in Ref.~\cite{Denicol:2012cn} that is, keeping only the second-order terms in both Knudsen numbers (gradients) and inverse Reynolds numbers (deviations from local equilibrium) in the evolution equations. We anticipate here that $\delta B^{\mu\nu}$, as necessary as it is for internal consistency, does not contribute to the second-order transport coefficients. Thanks to~(\ref{deltaB_evol}), it is at least second-order itself, and it can contribute only to third-order terms.

\section{Evolution equations for the dissipative currents}\label{sec:second_order}
\subsection{General approach}
%
%The novel feature of this work is to derive the second-order evolution equations for various dissipative currents within the system under consideration. In particular, the dissipative part of the baryon number current, i.e., the diffusion current $\nu^{\mu}$, the shear stress $\pi^{\mu\nu}$, and bulk pressure $\Pi$. To achieve this goal, we follow the computations reported at the beginning of~\cite{Tinti:2018qfb}. Calling ${\mathfrak f}^{\mu_1\cdots\mu_s}_r$ the moment of a generic distribution function $f$

In this work, we will extend the method used in Ref.~\cite{Tinti:2016bav} in the case of a strictly vanishing baryon current, $\mu=0$ and $J^\mu=0$. We will use the time derivatives of the physical tensors $\dot J^\mu =u\cdot \partial J^\mu$ and $\dot T^{\mu\nu} = u\cdot \partial T^{\mu\nu}$ as the sole basis for all the evolution equations. Then The projection of them with the four-velocity does not provide any new equations. They are, in fact, just the local conservation equations. They have to be used in hydrodynamics in any case and, conveniently, they do not contain any coupling with non-hydrodynamic degrees of freedom. The orthogonal projections, on the other hand, have such couplings and must be approximated\footnote{That is, on top of the relaxation type evolution of the generalized distribution functions, with a single relaxation time and the chosen form of the local equilibrium. Clearly, they also constitute an approximation and are not general enough to reproduce every kind of dynamics.} to obtain a closed set of equations in the hydrodynamic degrees of freedom only.

Proceeding in order, the fact that the $u_\mu$ projections of the $\dot J^\mu$ and $\dot T^{\mu\nu}$ end up giving back the conservation equations is not trivial. It can be seen using the general property
of such contractions
\begin{equation}\label{contr_prop}
    u_\mu \dot {\cal O}^{\mu\nu_1\cdots\nu_n} =  u_\mu u^\rho\partial_\rho {\cal O}^{\mu\nu_1\cdots\nu_n}= \partial_\mu{\cal O}^{\mu\nu_1\cdots\nu_n} -\nabla_\mu{\cal O}^{\mu\nu_1\cdots\nu_n},
\end{equation}
which is valid for any rank $n+1$ tensor ${\cal O}^{\mu\nu_1,\nu_n}$. Starting from the simplest case, the baryon current time derivative, calling $W^-$ the weight
\begin{equation}
    W^-(x,p) = \frac{g_q}{(2\pi^3)} 2 \Theta(p_0)\delta(p^2-M^2) f^-,
\end{equation}
one has 
\begin{equation}
    u_\mu \dot J^\mu = q\int d^4 p \left[ (p\cdot u) \, u\cdot\partial W^- \right] = q\int d^4 p \left[ p\cdot\partial W^- -p\cdot\nabla W^- \vphantom{\frac{}{}}\right] = -\nabla_\mu J^\mu -\frac{u_\mu}{\taueq}\left[ \vphantom{\frac{}{}} J^\mu - J^\mu_0\right] = -\nabla_\mu J^\mu.
\end{equation}
The last to last passage is done expanding the $p\cdot \partial W^-$, using Eq.~(\ref{delta_prime_prop}) and the dynamical equation~(\ref{Bol_Vlas_-}). The last passage is a consequence of the decomposition~(\ref{J_dec}), the local equilibrium form~(\ref{Jeq}), and the definition itself of the effective intensive parameters~(\ref{def_T_mu}). On the other hand, using also Eq.~(\ref{contr_prop})
\begin{equation}
    u_\mu\dot J^\mu = \partial_\mu J^\mu -\nabla_\mu J^\mu = -\nabla_\mu J^\mu, \quad \Rightarrow \quad \partial_\mu J^\mu=0.
\end{equation}
The situation is similar for $\dot T^{\mu\nu}$, and one can use a similar sequence of steps. The extra term in the $W^+$ weight
\begin{equation}
    W^+(x,p) = \frac{g}{(2\pi^3)} 2 \Theta(p_0)\delta(p^2-m^2) f + \frac{g_q}{(2\pi^3)} 2 \Theta(p_0)\delta(p^2-M^2) f^+,
\end{equation}
is hardly a problem, and $u_\mu \dot B^{\mu\nu}$ can be treated with~(\ref{contr_prop}) too. The main thing to remember is to treat properly the $-\frac{\partial}{\partial p_\mu}\delta(\cdots)$ from~(\ref{delta_prime_prop}). On top of the momentum derivatives of the generalized distribution functions, there is also the $g^{\mu\nu}$ from $\frac{\partial}{\partial p_\mu} p^\nu=g^\mu$ from the remaining momentum in the integral. Summing up
\begin{equation}
    \begin{split}
        \partial_\mu T^{\mu\nu} -\nabla_\mu T^{\mu\nu} &= u_\mu \dot T^{\mu\nu} = \int d^4 p \, p^\nu (p\cdot u) \dot W^+ \:+\; u_\mu \dot B^{\mu\nu}  \\
        &=-\nabla_\mu \int d^4p \, p^\mu p^\nu W^+ \;+\; \int d^4 p \, p\cdot \partial W^+ \;+\; \partial_\mu B^{\mu\nu} -\nabla_\mu B^{\mu\nu}\\
        &= -\nabla_\mu T^{\mu\nu} \;+\; \partial_\mu B^{\mu\nu} \;+\;\frac{u_\mu}{\taueq}\deltaB^{\mu\nu} \;+\; m\, \partial^\nu m\int_{\bf p} f \;+\; M\partial^\nu M \int_{\bf q} f^+ \\
        &= -\nabla_\mu T^{\mu\nu}. 
    \end{split}
\end{equation}
In the last passage, one recognizes the right-hand side of~(\ref{cons_B}), which must be zero to preserve the local conservation of four-momentum. The dynamical equations~(\ref{cons_deltaB}) coupled with~(\ref{thermocondition}) from the Gibbs-Duhem relations ensure that it vanishes. Therefore, also the $u_\mu \dot T^{\mu\nu}$ provide just conservation equations.

The projections orthogonal to $u_\mu$, namely $\dot J^{\langle\mu\rangle}$ and $\dot T^{\langle\mu\rangle\langle\nu\rangle}$, provide the time evolution of $\pi^{\mu\nu}$, $\Pi$ and $\nu^\mu$, since they contain all their time $u\cdot\partial$ derivatives. As long as one is interested only in second-order viscous hydrodynamics, the bag term disappears and one can consider only the kinetic-like terms. For the baryon current $\dot J^{\langle\mu\rangle}$ this is obvious, as there is no bag term in $J^\mu$. On the other hand, according to the decomposition~(\ref{T_dec}) one has
\begin{equation}
    \dot T^{\langle\mu\rangle\langle\nu\rangle} = \dot \pi^{\langle\mu\rangle\langle\nu\rangle} - \left( \dot \pres + \dot \Pi \right)\Delta^{\mu\nu},
\end{equation}
on the other hand, calling $\Tkin^{\mu\nu}$
\begin{equation}\label{T_kin}
    \begin{split}
        \Tkin^{\mu\nu} = \int_{\bf p} p^\mu p^\nu f +\int_{\bf q} p^\mu p^\nu \fp, \qquad \Tkinzero^{\mu\nu} =\int_{\bf p} p^\mu p^\nu \fn +\int_{\bf q} p^\mu p^\nu \fnp = \ped_0 \, u^\mu u^\nu - \pres_0 \Delta^{\mu\nu},
    \end{split}
\end{equation}
and $\Tkinzero^{\mu\nu}$ its local equilibrium expectation value, one has
\begin{equation}
\begin{split}
    \dot T^{\langle\mu\rangle\langle\nu\rangle} &=   \Tkindot^{\langle\mu\rangle\langle\nu\rangle} + \dot B^{\langle\mu\rangle\langle\nu\rangle} = \delta\Tkindot^{\langle\mu\rangle\langle\nu\rangle} + \Tkinzerodot^{\langle\mu\rangle\langle\nu\rangle} +\dot B_0\Delta^{\mu\nu}+ \delta\dot B^{\langle\mu\rangle\langle\nu\rangle} = \delta\Tkindot^{\langle\mu\rangle\langle\nu\rangle}+ \delta\dot B^{\langle\mu\rangle\langle\nu\rangle} -\dot \pres \Delta^{\mu\nu},\\
    &\Rightarrow \dot \pi^{\langle\mu\rangle\langle\nu\rangle} - \dot \Pi\, \Delta^{\mu\nu} = \delta\Tkindot^{\langle\mu\rangle\langle\nu\rangle}+ \delta\dot B^{\langle\mu\rangle\langle\nu\rangle}.
\end{split}
\end{equation}
The auxiliary $\delta \Tkin^{\mu\nu}$ is clearly the subtraction $\delta \Tkin^{\mu\nu}=\Tkin^{\mu\nu}-\Tkinzero^{\mu\nu}$. The evolution of the remaining five degrees of freedom of $T^{\mu\nu}$ depends only on $\delta \Tkin^{\mu\nu}$ and $\delta \dot B^{\mu\nu}$, but thanks to the evolution equations~(\ref{deltaB_evol}) this is at least third order. Therefore, up to higher-order terms, one has
\begin{equation}\label{pressure corrections_2nd_ord}
    \begin{split}
        \dot \pi^{\langle\mu\rangle\langle\nu\rangle} - \dot \Pi\, \Delta^{\mu\nu} & \simeq \delta\Tkindot^{\langle\mu\rangle\langle\nu\rangle} = \Tkindot^{\langle\mu\rangle\langle\nu\rangle} - \Tkinzerodot^{\langle\mu\rangle\langle\nu\rangle} = \Tkindot^{\langle\mu\rangle\langle\nu\rangle} + \dot \pres_0 \Delta^{\mu\nu},
    \end{split}
\end{equation}
or, equivalently, introducing the transverse traceless and symmetric projector $\Delta^{\mu\nu}_{\rho\sigma}$
\begin{equation}
    \Delta^{\mu\nu}_{\rho\sigma}= \frac{1}{2}\left( \vphantom{\frac{}{}} \Delta^\mu_\rho\Delta^\nu_\sigma + \Delta^\mu_\sigma\Delta^\nu_\rho -\frac{2}{3}\Delta^{\mu\nu}\Delta_{\rho\sigma}\right), \quad \longrightarrow {\cal O}^{\langle\mu\nu\rangle\alpha_1\cdots\alpha_n} = \Delta^{\mu\nu}{\cal O}^{\rho\sigma\alpha_1\cdots\alpha_n}_{\rho\sigma}
\end{equation}
and its contraction with a generic $n+2$-rank tensor, one has
\begin{equation}
    \dot \pi^{\langle\mu\rangle\langle\nu\rangle} = \Tkindot^{\langle\mu\nu\rangle}, \qquad \dot\Pi \simeq -\frac{1}{3} \Delta_{\mu\nu} \Tkindot^{{\langle\mu\rangle\langle\nu\rangle}} -\dot \pres_0.
\end{equation}
Since $\pres_0$ is a function of $\alpha$ and $\beta$ it is exactly known in terms of the intensive parameters and their derivatives
\begin{equation}\label{-P0dot}
    \begin{split}
        -\dot\pres_0 &= -\dot m\, \frac{\partial \pres_0}{\partial m} -\dot M\, \frac{\partial \pres_0}{\partial M} -\dot \alpha \left[ \frac{\partial \pres_0}{\partial T}\frac{\partial T}{\partial\alpha} + \frac{\partial \pres_0}{\partial \mu}\frac{\partial \mu}{\partial\alpha}  \right] -\dot \beta \left[ \frac{\partial \pres_0}{\partial T}\frac{\partial T}{\partial\beta} + \frac{\partial \pres_0}{\partial \mu}\frac{\partial \mu}{\partial\beta}  \right] \\
        &= m\, \dot m \int_{\bf p} \fn + M\, \dot M\int_{\bf q}\fnp -\frac{\rho}{\beta} \, \dot \alpha +\frac{\ped+\pres}{\beta}\, \dot\beta \\
        &=\left[ m\frac{\partial m}{\partial\alpha}\int_{\bf p}\fn + M\frac{\partial M}{\partial\alpha}\int_{\bf q}\fnp -\frac{\rho}{\beta} \right] \dot \alpha + \left[ m\frac{\partial m}{\partial\beta}\int_{\bf p}\fn + M\frac{\partial M}{\partial\beta}\int_{\bf q}\fnp +\frac{\ped +\pres}{\beta} \right] \dot \alpha.
     \end{split}
\end{equation}
The terms in the square brackets are of zeroth order. The $\dot \alpha$ and $\dot \beta$ have been already computed in~(\ref{alpha_beta_dot}), and they are just second order. In the end, only $\dot\Tkin^{\langle\mu\rangle\langle\nu\rangle}$ and $\dot J^{\langle\mu\rangle}$ have to be expanded and properly approximated to second order. Conveniently enough, since they are all kinetic-like terms, they can be treated with the methods already used for the Boltmzann-Vlasov equation. Calling ${\mathfrak f}^{\mu_1\cdots\mu_s}_r$ the rank $s$ tensorial moment
\begin{equation}
   {\mathfrak f}^{\mu_1\cdots\mu_s}_r= \Delta^{\mu_1}_{\nu_1}\cdots \Delta^{\mu_s}_{\nu_s} \int_{\bf p} (p\cdot u)^r \, p^{\nu_1}\cdots p^{\nu_s} f,
\end{equation}
and the analogous definitions for the $f^\pm$ integrals. It is possible to use the exact evolution equation found in Ref.~\cite{Tinti:2018qfb} or, more precisely, its restriction to this case: medium dependent mass $m$, not Lorentz Force and collisions of the relaxation type. In other words
\begin{equation}\label{exact_dot_frakf}
    \begin{split}
         \dot{\mathfrak f}^{\langle\mu_1\rangle\cdots\langle\mu_s\rangle}_r +\frac{1}{\tau_q}\delta {\mathfrak f}^{\langle\mu_1\rangle\cdots\langle\mu_s\rangle}_r&=
          (r-1)m\, \dot m \; {\mathfrak f}^{\mu_1\cdots\mu_s}_{r-2}+ r\; \dot u_\nu{\mathfrak f}^{\nu\mu_1\cdots\mu_s}_{r-1} - s \; \dot u^{(\mu_1}{\mathfrak f}^{\mu_2\cdots\mu_s)}_{r+1} \\&+ s\;  m \nabla^{(\mu_1}m \; {\mathfrak f}^{\mu_2\cdots\mu_s)}_{r-1}-\theta\, {\mathfrak f}^{\mu_1\cdots\mu_s}_{r}   -\nabla_\nu {\mathfrak f}^{\nu \langle\mu_1\rangle\cdots\langle\mu_s\rangle}_{r-1}\\
         &\qquad -s\; \nabla_\nu u^{(\mu_1}{\mathfrak f}^{\mu_2\cdots\mu_s)\nu}_{r} +(r-1)\; \nabla_\nu u_\lambda \;  {\mathfrak f}^{\nu\lambda\mu_1\cdots\mu_s}_{r-2},
    \end{split}
\end{equation}
and the corresponding formulas for the $f^\pm$ case with $M$ instead of $m$. As long as one considers only second-order viscous hydrodynamics, no problem with the infrared divergences mentioned in\,~\cite{Tinti:2018qfb} appears if either of the masses vanishes. The negative $r$ in the equations for $ \Tkindot^{\langle\mu\rangle\langle\nu\rangle}$ and $\dot J^{\langle\mu\rangle}$ are never small enough to provide non-integrable poles in the right-hand side. If one wants to extend the method either tensor moments must be considered, for instance, the resummed ones in the reference itself~\cite{Tinti:2018qfb}.

It is convenient for this work to rewrite the above evolution equations~(\ref{exact_dot_frakf}). They explicitly state the dependence on the tensorial $\mathfrak f$ moments, but for some computations, it is more convenient the equivalent form
\begin{equation}\label{more_conventient_exact_dot_frakf}
    \begin{split}
         \dot{\mathfrak f}^{\langle\mu_1\rangle\cdots\langle\mu_s\rangle}_r +\frac{1}{\taueq}\delta {\mathfrak f}^{\langle\mu_1\rangle\cdots\langle\mu_s\rangle}_r
         &=  (r-1)\, m\,\dot m \,{\mathfrak f}_{r-2}^{\mu_1\cdots\mu_s} +r \; \dot u_\nu \, {\mathfrak f}^{\nu\mu_1\cdots\mu_s}_{r-1}- s \; \dot u^{(\mu_1}{\mathfrak f}^{\mu_2\cdots\mu_s)}_{r+1} \\
         & \qquad - \int_{\bf p} (p\cdot u)^{r-1} p^{\langle\mu_1\rangle}\cdots p^{\langle\mu_s\rangle}\left[ \vphantom{\frac{}{}} p\cdot\nabla f +m\nabla_\nu m \partial^\nu_{(p)}f\right].
    \end{split}
\end{equation}
%
%Henceforth, whether utilizing Eq.~\eqref{exact_dot_frakf} or~\eqref{more_conventient_exact_dot_frakf} will give equivalent forms of the evolution of the various dissipative currents. 
The method that will be used to obtain a closed set of equations for second-order viscous hydrodynamics, is an extension of the method used in Ref.~\ref{Tevolution}. Namely, to the case of three generalized distribution functions with two different medium-dependent masses. In turn, that was a generalization of the method in Ref~\cite{Jaiswal:2013npa} to the case of a medium-dependent mass.

The main idea is, first, to write all the terms with their explicit dependence on the degrees of freedom. Then, substitute $f$ and $f^\pm$ in the others with their local equilibrium counterparts $\fn$ and $\fnpm$ to obtain the first-order approximations. That is cutting all the terms second order or higher, including the time derivative on the left-hand side. Then using the form of the equations~(\ref{Bol_Vlas_-}) and~(\ref{Bol_vlas_0+}) to make a first-order expansion in the gradients. Substituting explicit gradients with the first-order approximations, and then plugging back the first-order approximation of the generalized distribution functions in the non-hydrodynamic tensor moment. Keeping only the terms up to the second order, this is the proposed second-order approximation.

This method to extract second-order approximations is not the only one. We will use it because, in the presence of an exact solution of the microscopic theory, the prescription in Ref.~\cite{Jaiswal:2013npa} has been rather accurate, even for large gradients and deviations from local equilibrium. It is fair to say, however, that in the absence of exact solutions, it is not possible to confirm that it will keep being such a good approximation.

%***************************
\subsection{Diffusion current}
%***************************
%
The starting equation for the diffusion current $\nu^\mu$ is the multiplication by $q$ of~\eqref{exact_dot_frakf}, specifically with $r=0$ and $s=1$. Of course, the distribution for the diffusion current must be $f^-$, not $f$, and the mass $M$. One can see that from
\begin{equation}
    \nu^\mu = J^{\langle\mu\rangle} = q\Delta^\mu_\rho \int_{\bf q} p^\rho f^-, \qquad \dot J^{\langle\mu\rangle} = \Delta^\mu_\rho\Delta^\rho\sigma \dot J^\rho =  \Delta^\mu_\rho\left[ \dot \nu^\rho + \rho\dot u^\mu \vphantom{\frac{}{}} \right] = \dot \nu^{\langle\mu\rangle} +\rho \dot u^\mu.
\end{equation}
Then, using the aforementioned version of~\eqref{exact_dot_frakf}
\begin{equation}\label{exact_dot_nu}
    \begin{split}
         \dot \nu^{\langle\mu\rangle} +\frac{1}{\taueq} \, \nu^\mu  &=-q\; M\, \dot M \; \int_\bq \frac{p^{\langle\mu\rangle}}{(p\cdot u)^2}f^- + q\; M \nabla^{\mu}M \int_\bq \frac{f^-}{(p\cdot u)} \; -\theta\, \nu^\mu - \dot u^{\mu} \, \rho -q\, \Delta^\mu_\lambda\nabla_\nu \int_\bq \frac{p^{\langle\nu\rangle}p^{\langle\lambda\rangle}}{(p\cdot u)}f^-\\
         &\qquad -(\nu\cdot \partial) u^{\mu}- q\; \nabla_\nu u_\lambda \int_\bq \frac{p^{\langle\nu\rangle}p^{\langle\lambda\rangle}p^{\langle\mu\rangle}}{(p\cdot u)^2}f^-,
    \end{split}
\end{equation}
This is the most convenient form to use to obtain the second-order approximation. For the first order one, it is faster to use an equivalent approach, starting from~\eqref{more_conventient_exact_dot_frakf} instead, the evolution of $\nu^{\mu}$ then reads 
\begin{equation}\label{convenient_exact_dot_nu}
    \begin{split}
         \dot \nu^{\langle\mu\rangle} +\frac{1}{\taueq} \, \nu^\mu  &= -\rho \, \dot u^\mu -M\,\dot M \; q\!\int_\bq\frac{p^{\langle\mu\rangle}}{(p\cdot u)^2}f^- - q\!\int_\bq \frac{p^{\langle\mu\rangle}}{(p\cdot u)}\left[ \vphantom{\frac{}{}} p\cdot\nabla f^- +M\nabla_\nu M\,  \partial^\nu_{(p)}f^-\right].
    \end{split}
\end{equation}
The time derivative itself of $\nu^\mu$ in the left-hand side of Eq.~\eqref{convenient_exact_dot_nu} drops, being second order. Concerning the right-hand side, it takes a few more steps to have a convenient form. The first term there is already first order, and the other ones are obtained by substituting $f^-$ with $\fnm$. One can already see that the second term on the right-hand side of~\eqref{convenient_exact_dot_nu} is vanishing since $\fnm$ is rotation invariant in the comoving frame. The only non vanishing vector moments of $\fnm$ are proportional to $u^\mu$, but $u^{\langle\mu\rangle}=0$. The very last term is also vanishing, since $\partial^\nu_{(p)}\fnm\propto u^\mu$ and it is thus orthogonal to $\nabla_\nu M$. The gradient of $\fnm$ on the other hand reads
\begin{equation}
    p\cdot\nabla\fnm = q\left(p\cdot\nabla\alpha \vphantom{\frac{}{}}\right) \fnp -  (p\cdot u)\left(p\cdot \nabla \beta\vphantom{\frac{}{}}\right) \fnm -\beta \left(p^\rho p^\sigma \nabla_\rho u_\sigma \vphantom{\frac{}{}}\right)\fnm. 
\end{equation}
The last term does not contribute to the integral, since it adds two momenta orthogonal to $u^\mu$. The integral then reads
\begin{equation}\label{int_grad_f0-}
   - q\!\int_\bq \frac{p^{\langle\mu\rangle}}{(p\cdot u)}\left[ \vphantom{\frac{}{}} p\cdot\nabla \fnm \right]
   = q^2 \, I^+_{1,1} \, \nabla^\mu \alpha -\frac{\rho}{\beta}\, \nabla^\mu\beta
\end{equation}
The thermodynamic integrals $I_{n,m}^\pm$ and $I_{n,m}$ are
\begin{equation}
    I^\pm_{m,n}=\frac{1}{(2n+1)!!}\int_{\bf p}(p\cdot u)^{m-2n}(-p\cdot\Delta\cdot p)^n \, \fn^\pm, \qquad I_{m,n}=\frac{1}{(2n+1)!!}\int_{\bf p}(p\cdot u)^{m-2n}(-p\cdot\Delta\cdot p)^n \, \fn.
\end{equation}
In some cases, they can be written as simple expressions of familiar quantities, like in the second term on the right-hand side of~\eqref{int_grad_f0-}, but in most cases, they will be left as they appear canonically in the transport coefficients.
In order to keep the first-order terms only, we make use of the general formula~\eqref{alpha-beta_grad}. the integral~\eqref{int_grad_f0-}, then reads
\begin{equation}
    - q\!\int_\bq \frac{p^{\langle\mu\rangle}}{(p\cdot u)}\left[ \vphantom{\frac{}{}} p\cdot\nabla \fnm \right]
   = \left[ q^2 \, I^+_{1,1} -\frac{\rho^2}{\beta(\ped+\pres)}\right]\, \nabla^\mu \alpha +\rho\left( 1+\frac{\Pi}{\ped+\pres} \right)\, \dot u^\mu -\frac{\rho}{\ped+\pres} \left(  \vphantom{\frac{}{}} \nabla^\mu \Pi-\Delta^\mu_\rho\, \partial_\sigma \pi^{\rho\sigma}\right).
\end{equation}
Keeping only the first order terms, the contribution from the acceleration $\dot u^\mu$ simplifies exactly, while it remains
\begin{equation}\label{diff_first}
    \nu^\mu \simeq\kappa_b\,\nabla^\mu\alpha. %\qquad\left[ \vphantom{\frac{}{}}\mbox{plus higher order terms}\right]
\end{equation}
The diffusion current, as expected, is proportional to the gradient of the effective $\alpha =\mu/T$. That is the effective parameter controlling the baryon charge density\footnote{Te sign set is the density is positive or negative, and the magnitude, coupled to $\beta=1/T$ itself, selects the ratio of $\rho/\ped$.}. The (single) first-order transport coefficient for $\nu^\mu$ reads
\begin{equation}\label{diff_first_transportt}
     \kappa_{b}=\taueq \left( \vphantom{\frac{}{}}  q^2\, I_{1,1}^+ \; - \; \frac{\rho^2}{\beta(\ped+\pres)} \right).
\end{equation}
The remaining term
\begin{equation}
    \frac{\rho }{\ped+\pres} \, \Pi\,  \dot u^\mu -\frac{\rho}{\ped+\pres} \left(  \vphantom{\frac{}{}} \nabla^\mu \Pi-\Delta^\mu_\rho\, \partial_\sigma \pi^{\rho\sigma}\right),
\end{equation}
represents the only second-order (or higher) contribution to the evolution equation from $f_0^-$. It must be included in the second-order equation, but thanks to this analysis, all the remaining second-order terms must come from $\delta f^-$. because of this, to compute the full second-order approximation, it is convenient to switch to the form~(\ref{exact_dot_nu}). Making use of the previous results
\begin{equation}\label{almpst_finished_nu}
    \begin{split}
         \dot \nu^{\langle\mu\rangle} +\frac{1}{\taueq} \, \nu^\mu  =& \frac{\kappa_b}{\taueq} \, \nabla^\mu\alpha +\frac{\rho }{\ped+\pres} \, \Pi\,  \dot u^\mu -\frac{\rho}{\ped+\pres} \left(  \vphantom{\frac{}{}} \nabla^\mu \Pi -\Delta^\mu_\rho\, \partial_\sigma \pi^{\rho\sigma}\right) \\
         &-\; M\, \dot M \; q\,\int_\bq \frac{p^{\langle\mu\rangle}}{(p\cdot u)^2}\delta  f^-_{1} + q\; M \nabla^{\mu}M \int_\bq \frac{ \delta  f^-_{1}}{(p\cdot u)} \; -\theta\, \nu^\mu  \\
         & -q\, \Delta^\mu_\lambda\nabla_\nu \int_\bq \frac{p^{\langle\nu\rangle}p^{\langle\lambda\rangle}}{(p\cdot u)}\delta  f^-_{1}-(\nu\cdot \partial) u^{\mu}- q\; \nabla_\nu u_\lambda \int_\bq \frac{p^{\langle\nu\rangle}p^{\langle\lambda\rangle}p^{\langle\mu\rangle}}{(p\cdot u)^2}\delta f^-_{1}.
    \end{split}
\end{equation}
One then needs just the first-order approximation of $f^-$, or rather $\delta f^-_{1} = f^-_{1}-f^-_0$, and keep just the second order terms. Starting from the Eqs.~\eqref{Bol_vlas_qqbar} or, equivalently, their sum and subtraction in~\eqref{Bol_vlas_0+} and~\eqref{Bol_Vlas_-} one has
\begin{equation}\label{grad_fpm_def}
    p\cdot f^\pm +M\, \partial_\mu M \, \partial^\mu_{(p)}f^\pm =-\frac{(p\cdot u)}{\taueq}\left( \vphantom{\frac{}{}} f^\pm -\fnpm \right)\quad \Rightarrow\quad f^\pm = \fnpm -\frac{\taueq}{(p\cdot u)}\left[ p\cdot \partial f^\pm +M\, \partial_\mu M \, \frac{\partial f^\pm}{\partial_{p_\mu}}\right].
\end{equation}
The local equilibrium $f^\pm_0$ is clearly the zeroth-order approximation. For the first order approximation, one starts plugging $f^\pm_0$ in the right hand side of~\eqref{grad_fpm_def}
\begin{equation}\label{deltaf_1_0}
    \begin{split}
        f^\pm_1 \simeq& \fnpm -\frac{\taueq}{(p\cdot u)}\left[ p\cdot \partial \, \fnpm +M\, \partial_\mu M \, \frac{\partial \fnpm}{\partial_{p_\mu}}\right] \quad \Rightarrow \quad \delta f^\pm_1 \simeq  -\frac{\taueq}{(p\cdot u)}\left[ p\cdot \partial \, \fnpm \;-\;\beta \, M\dot M \, \fnpm \vphantom{\frac{}{}}\right].
    \end{split}
\end{equation}
Both $\fnpm$ and $M$ depend on $\alpha$ and $\beta$. As seen in the previous steps, their gradients contain both first-order and second-order terms. In order to have a proper first-order correction, the latter must be removed. In order to do so, we expand explicitly the the gradients. Starting from the simpler term in the $M$ derivatives
\begin{equation}\label{Mp.dM}
    \begin{split}
        M \dot M &=  M \left[ \frac{\partial M}{\partial \alpha} \, \dot \alpha \;+\; \frac{\partial M}{\partial \beta} \, \dot \beta \right] \simeq M \left[ \frac{\partial M}{\partial \alpha} \, D_{(\alpha)} \;+\; \frac{\partial M}{\partial \beta} \, D_{(\beta)} \right] \; \theta,
    \end{split}
\end{equation}
in which we made use of the first-order approximations of the exact ~\eqref{alpha_beta_dot}
\begin{align}
    %\nabla^\mu \beta & \simeq \frac{\rho}{\ped+\pres} \, \nabla^\mu \alpha - \beta \, \dot u^\mu, \\ \nonumber \\
    \dot \alpha &\simeq \left[\frac{\rho \, \tfrac{\partial \, \pedeq}{\partial\beta} - (\ped +\pres)\tfrac{\partial \rhoeq}{\partial\beta}}{\tfrac{\partial \, \pedeq}{\partial\alpha} \, \tfrac{\partial \, \rhoeq}{\partial\beta}-\tfrac{\partial \, \pedeq}{\partial\beta} \, \tfrac{\partial \, \rhoeq}{\partial\alpha}} \right] \; \theta \; = \; D_{(\alpha)}\;  \theta, \\
    \dot \beta &\simeq \left[\frac{(\ped +\pres) \, \tfrac{\partial \, \rhoeq}{\partial\alpha} - \rho\tfrac{\partial \pedeq}{\partial\alpha}}{\tfrac{\partial \, \pedeq}{\partial\alpha} \, \tfrac{\partial \, \rhoeq}{\partial\beta}-\tfrac{\partial \, \pedeq}{\partial\beta} \, \tfrac{\partial \, \rhoeq}{\partial\alpha}} \right] \; \theta \; = \; D_{(\beta)}\;  \theta,
\end{align}
and the relative $D_{(\alpha)}$, $D_{(\beta)}$ short-hand notation. The gradient in $\fnpm$ is rather more complicated
\begin{equation}
    \begin{split}
        p\cdot\partial\, \fnpm =& q\,(p\cdot\partial\, \alpha) \fnmp -\left[(p\cdot u)(p\cdot\partial \, \beta) \;+\; \beta \, p^\mu p^\nu \, \partial_\mu u_\nu \vphantom{\frac{}{}}\right] \fnpm \\
        =& q\left[ (p\cdot u) \, \dot \alpha \;+\; p\cdot \nabla \alpha \vphantom{\frac{}{}}\right] \fnmp \\
        &-\left[(p\cdot u)^2\, \dot \beta +(p\cdot u)(p\cdot \nabla \beta) + \beta(p\cdot u)(p\cdot \dot u) +\beta\left( p\cdot \sigma\cdot p +\frac{1}{3}\theta \, p\cdot \Delta\cdot p\right) \right] \fnpm.
    \end{split}
\end{equation}
Making use of the first-order approximations~\eqref{Mp.dM} and the first order terms only from~\eqref{alpha-beta_grad}, the gradient reads
\begin{equation}
    \begin{split}
        p\cdot\partial\, \fnpm \simeq&  \quad q\left[ (p\cdot u) D_{( \alpha)} \, \theta \;+\; p\cdot \nabla \alpha \vphantom{\frac{}{}}\right] \fnmp \\
        &-\left[(p\cdot u)^2D_{(\beta)} \, \theta +(p\cdot u)\,\frac{\rho}{\ped+\pres} \, (p\cdot \nabla \alpha)+\beta\left( p\cdot \sigma\cdot p +\frac{1}{3}\theta \, p\cdot \Delta\cdot p\right) \right] \fnpm.
    \end{split}
\end{equation}
The first order approximation $\delta f^\pm_1$ would then read
\begin{equation}
    \begin{split}
        \delta\fpm_1 \simeq& -q\left[ \taueq \, D_{(\alpha)} \, \theta +\frac{\taueq}{(p\cdot u)}(p\cdot\nabla\alpha) \right]\fnmp \\
        &+\left[ \taueq (p\cdot u)D_{(\beta)} \, \theta +\frac{\taueq \, \rho}{\ped+\pres} \, (p\cdot \nabla \alpha)+\frac{\taueq\,\beta}{(p\cdot u)}\left( p\cdot \sigma\cdot p +\frac{1}{3}\theta \, p\cdot \Delta\cdot p\right)\right]\fnpm \\
        & +\frac{\taueq \, \beta \, M}{(p\cdot u)}\left( \frac{\partial M}{\partial \alpha} \, D_{(\alpha)} \;+\; \frac{\partial M}{\partial \beta} \, D_{(\beta)} \right) \fnmp.
    \end{split}
\end{equation}
Since the multiplication of two gradients is inconvenient for numerical purposes (it tends to make the system mathematically unstable), we opt to use another form for $\delta f^\pm_1$. We substitute the gradients with their corresponding hydrodynamic degrees of freedom
\begin{equation}
    \nabla^\mu \alpha\simeq \frac{\nu^\mu}{\kappa_b}, \qquad \theta \simeq-\frac{\Pi}{\zeta}, \qquad \sigma^{\mu\nu}\simeq \frac{\pi^{\mu\nu}}{2\eta}. 
\end{equation}
Since the difference between the two is second-order or higher, it is still a first-order approximation. In the next section, the formulas for the $\eta$ and $\zeta$ will be presented. The first order $\delta f_1^\pm$ we will use is then
\begin{equation}\label{delta_f_pm_1}
    \begin{split}
        \delta f^\pm_{1} = & -\frac{\taueq}{\kappa_b}\left[  \frac{q}{(p\cdot u)} \fnmp- \frac{\rho}{(\ped+\pres)}\fnpm \right] (p\cdot \nu) +\frac{\taueq\, \beta}{2\eta(p\cdot u)} \, (p\cdot \pi\cdot p) \fnpm  \\
    & + \frac{\taueq \, \Pi}{\zeta}\left[ q\, D_{(\alpha)} \, \fnmp-(p\cdot u)\, D_{(\beta)} \, \fnpm +\frac{\beta\, (-p\cdot\Delta\cdot p)}{3(p\cdot u)} \, \fnpm -\frac{\beta\,  M}{(p\cdot u)} \left( \frac{\partial M}{\partial \alpha} D_{(\alpha)} + \frac{\partial M}{\partial \beta} D_{(\beta)} \right)\fnpm\right].
    \end{split}
\end{equation}
Therefore, plugging~\eqref{delta_f_pm_1} in~\eqref{almpst_finished_nu}, one obtains the second-order evolution equation of the diffusion current $\nu^\mu$  
\begin{align}
    \dot \nu^{\langle\mu\rangle} +\frac{1}{\taueq} \, \nu^\mu  =   &\frac{\kappa_b}{\taueq} \, \nabla^\mu\alpha \;+\;\tau_{\nu\Pi}\,\Pi \,\dot{u}^{\mu} \;-\; c_{\pi\Pi}\left( \nabla^\mu \Pi -\Delta^\mu_\rho \, \partial_\sigma \pi^{\rho\sigma} \vphantom{\frac{}{}}\right) \;+\;\delta_{\nu\nu}\, \theta\, \nu^{\mu} \;+\; c_{\nu\Pi}\, \Pi\, \nabla^\mu\alpha \\
    & \;-\; \Delta^\mu_\lambda \nabla_\rho \left[ l_{\nu\Pi}\, \Pi\, \Delta^{\mu\nu} -l_{\nu\pi} \, \pi^{\mu\nu} \vphantom{\frac{}{}}\right]\;+\;\lambda_{\nu\nu}\, \sigma^{\mu\lambda}\, \nu_{\lambda} \;+\;\omega^{\mu\lambda}\nu_{\lambda}.\label{diffusion_compact}
\end{align}
The explicit formulas for the second-order transport coefficients are listed in Appendix~\eqref{app_tc}.

\subsection{Shear stress and bulk pressure}

%***************************
In this section, we obtain the evolution equations of the pressure corrections $\pi^{\mu\nu}$ and $\Pi$ at second the second order. The procedure is essentially the same as the one seen in the last chapter for the diffusion current. Thanks to Eq.\eqref{pressure corrections_2nd_ord}, it is sufficient to compute $\Tkindot^{\mu\nu}$, that is,  applying the formulas in~(\ref{exact_dot_frakf}) with $r=0$ and $s=2$ for $f^+$ and $f$
\begin{equation}\label{Tevolution}
    \begin{split}
\Tkindot^{\langle\mu\rangle\langle\nu\rangle} +\frac{1}{\taueq} \delta \Tkin^{\langle\mu\rangle\langle\nu\rangle} \simeq &-m\,\dot m \int_{\bf p} \frac{p^{\langle\mu\rangle}p^{\langle\nu\rangle}}{(p\cdot u)^2}f -M\dot M \int_{\bf q} \frac{p^{\langle\mu\rangle}p^{\langle\nu\rangle}}{(p\cdot u)^2}f^+ + 2 m\nabla^{(\mu}m\int_{\bf p}\frac{p^{\langle\nu\rangle)}}{(p\cdot u)}f + 2 M\nabla^{(\mu}M\int_{\bf q}\frac{p^{\langle\nu\rangle)}}{(p\cdot u)}f^+ \\
        &-\Delta^\mu_\alpha \Delta^\nu_\beta\nabla_\lambda\left[ \int_{\bf p} \frac{p^{\langle\lambda\rangle}p^{\langle\alpha\rangle}p^{\langle\beta\rangle}}{(p\cdot u)}f +\int_{\bf q} \frac{p^{\langle\lambda\rangle}p^{\langle\alpha\rangle}p^{\langle\beta\rangle}}{(p\cdot u)}f^+  \right] \\
        & -\nabla_\lambda u_\sigma \left[ \int_{\bf p} \frac{p^{\langle\lambda\rangle}p^{\langle\sigma\rangle}p^{\langle\mu\rangle}p^{\langle\nu\rangle}}{(p\cdot u)^2}f +\int_{\bf q} \frac{p^{\langle\lambda\rangle}p^{\langle\sigma\rangle}p^{\langle\mu\rangle}p^{\langle\nu\rangle}}{(p\cdot u)^2}f^+ \right] \\
        &+ \theta\left( \vphantom{\frac{}{}} \pres +\Pi \right) \Delta^{\mu\nu} - \theta\pi^{\mu\nu} +2\left( \vphantom{\frac{}{}} \pres +\Pi \right) \nabla^{(\mu}u^{\nu)} - 2\nabla_\lambda u^{(\mu}\pi^{\nu)\lambda},
    \end{split}
\end{equation}
the terms on the right-hand side that would have been third-order or higher have been omitted since they do not contribute.

Also, from~\eqref{pressure corrections_2nd_ord} and~\eqref{Bnoneq}
\begin{equation}
\begin{split}
&\dot\pi^{\langle\mu\rangle\langle\nu\rangle} -\dot\Pi\Delta^{\mu\nu} = \Tkindot^{\langle\mu\rangle\langle\nu\rangle}+\dot \pres_0\Delta^{\mu\nu} \; \longrightarrow \\
    &\longrightarrow \;\dot\pi^{\langle\mu\rangle\langle\nu\rangle} -\dot\Pi\Delta^{\mu\nu} +\frac{1}{\taueq}\left( \pi^{\mu\nu} -\Pi\Delta^{\mu\nu} \vphantom{\frac{}{}}\right) = \Tkindot^{\langle\mu\rangle\langle\nu\rangle} +\frac{1}{\taueq}\delta\Tkin^{\mu\nu} +\frac{1}{\taueq}\, b_0\,\Delta^{\mu\nu}+\dot \pres_0\Delta^{\mu\nu},
    \end{split}
\end{equation}
making use of~\eqref{delta_f_both} and~\eqref{-P0dot}, and keeping up to the second-order terms only
\begin{equation}\label{mod_pres_corr_ev}
\begin{split}
&\dot\pi^{\langle\mu\rangle\langle\nu\rangle} -\dot\Pi\Delta^{\mu\nu} +\frac{1}{\taueq}\left( \pi^{\mu\nu} -\Pi\Delta^{\mu\nu} \vphantom{\frac{}{}}\right) \\
    & =\Tkindot^{\langle\mu\rangle\langle\nu\rangle} +\frac{1}{\taueq}\delta\Tkin^{\mu\nu} -\left[ n\dot m \int_{\bf p} f +M\dot M\int_{\bf q}f^+ +\frac{(\ped+\pres)}{\beta}\, \dot \beta -\frac{\rho}{\beta} \, \dot \alpha\right]\Delta^{\mu\nu},
    \end{split}
\end{equation}
the $\dot \alpha$ and $\dot \beta$ are already strictly of second order thanks to~\eqref{alpha_beta_dot}. One must keep only the second-order terms from the $\dot m$ and $\dot M$ parts. Adding the last term on the right-hand side of~\eqref{mod_pres_corr_ev} to~\eqref{Tevolution}, one has
\begin{equation}\label{Tevolution_1}
    \begin{split}
    \dot\pi^{\langle\mu\rangle\langle\nu\rangle} &-\dot\Pi\Delta^{\mu\nu} +\frac{1}{\taueq}\left( \pi^{\mu\nu} -\Pi\Delta^{\mu\nu} \vphantom{\frac{}{}}\right) \\
       \simeq&-\left[ m\dot m \int_{\bf p} f +M\dot M\int_{\bf q}f^+ +\frac{(\ped+\pres)}{\beta}\, \dot \beta -\frac{\rho}{\beta} \, \dot \alpha\right]\Delta^{\mu\nu}\\
       &-m\,\dot m \int_{\bf p} \frac{p^{\langle\mu\rangle}p^{\langle\nu\rangle}}{(p\cdot u)^2}f -M\dot M \int_{\bf q} \frac{p^{\langle\mu\rangle}p^{\langle\nu\rangle}}{(p\cdot u)^2}f^+ + 2 m\nabla^{(\mu}m\int_{\bf p}\frac{p^{\langle\nu\rangle)}}{(p\cdot u)}f + 2 M\nabla^{(\mu}M\int_{\bf q}\frac{p^{\langle\nu\rangle)}}{(p\cdot u)}f^+ \\
        &-\Delta^\mu_\alpha \Delta^\nu_\beta\nabla_\lambda\left[ \int_{\bf p} \frac{p^{\langle\lambda\rangle}p^{\langle\alpha\rangle}p^{\langle\beta\rangle}}{(p\cdot u)}f +\int_{\bf q} \frac{p^{\langle\lambda\rangle}p^{\langle\alpha\rangle}p^{\langle\beta\rangle}}{(p\cdot u)}f^+  \right] \\
        & -\nabla_\lambda u_\sigma \left[ \int_{\bf p} \frac{p^{\langle\lambda\rangle}p^{\langle\sigma\rangle}p^{\langle\mu\rangle}p^{\langle\nu\rangle}}{(p\cdot u)^2}f +\int_{\bf q} \frac{p^{\langle\lambda\rangle}p^{\langle\sigma\rangle}p^{\langle\mu\rangle}p^{\langle\nu\rangle}}{(p\cdot u)^2}f^+ \right] \\
        &+ \theta\left( \vphantom{\frac{}{}} \pres +\Pi \right) \Delta^{\mu\nu} - \theta\pi^{\mu\nu} +2\left( \vphantom{\frac{}{}} \pres +\Pi \right) \nabla^{(\mu}u^{\nu)} - 2\nabla_\lambda u^{(\mu}\pi^{\nu)\lambda},
    \end{split}
\end{equation}
the traceless part is the evolution of $\pi^{\mu\nu}$, while the contraction with $-\Delta_{\mu\nu}/3$ is the evolution for $\Pi$.

Following the same procedure as in the previous section, from the first-order terms only, we obtain from~\eqref{Tevolution_1}  the Navier-Stokes forms of the shear stress and bulk pressure,
\begin{align}\label{first-order-equations}
    \pi^{\mu\nu}\simeq 2 \eta\sigma^{\mu\nu}~,~~~\Pi\simeq -\zeta\theta,
\end{align}
the corresponding transport coefficients are
\begin{align}\label{1st-order-transport}
        & \eta = \, \taueq\left( \vphantom{\frac{}{}} \pres - I^+_{2,2} - I_{2,2} \right),\\
        &\zeta=\frac{5}{3}\eta  - \taueq\left[m\left( \frac{\partial m}{\partial\alpha}D_{(\alpha)}+ \frac{\partial m}{\partial\beta}D_{(\beta)} \right)\vphantom{\frac{}{}}\left( I_{0,1} -I_{0,0} \vphantom{I^+_{0,0}}\right) \right. \nonumber\\ 
        ~~~&~~~~\left. -  M\left( \frac{\partial M}{\partial\alpha}D_{(\alpha)}+ \frac{\partial M}{\partial\beta}D_{(\beta)} \right)\left(I_{0,1}^+-I^+_{0,0} \right)-\frac{(\ped+\pres)}{\beta} D_{(\beta)} +\frac{\rho}{\beta}D_{(\alpha)} \right].
\end{align}
The second-order approximation~\eqref{Tevolution_1} then reads, highlighting explicitly the first-order terms and making use of the exact formula~\eqref{alpha_beta_dot} with the short-hand notation~\eqref{short-hand_partials}
\begin{equation}\label{Tevolution_2}
    \begin{split}
    \dot\pi^{\langle\mu\rangle\langle\nu\rangle} &-\dot\Pi\Delta^{\mu\nu} +\frac{1}{\taueq}\left( \pi^{\mu\nu} -\Pi\Delta^{\mu\nu} \vphantom{\frac{}{}}\right) \simeq \frac{2\, \eta}{\taueq}\, \sigma^{\mu\nu} + \frac{\zeta}{\taueq} \, \theta\\
       &-\theta \left[ m\left( \frac{\partial m}{\partial\alpha} D_{(\alpha)} + \frac{\partial m}{\partial\beta} D_{(\beta)} \right) \int_{\bf p} \delta f_1 +M\left( \frac{\partial M}{\partial\alpha} D_{(\alpha)} + \frac{\partial M}{\partial\beta} D_{(\beta)} \right)\int_{\bf q}\delta f^+_1 \right] \Delta^{\mu\nu} \\
       &-\left[ \frac{(\ped+\pres)}{\beta}\, \frac{(\partial\cdot \nu)\, \ped_\beta -(\theta \Pi -\sigma:\pi)\, \rho_\beta}{{\cal J}} -\frac{\rho}{\beta} \, \frac{(\theta \Pi -\sigma:\pi)\, \rho_\alpha-(\partial\cdot \nu)\, \ped_\alpha }{{\cal J}} \right]\Delta^{\mu\nu}\\
       &- \theta\, m\,\left( \frac{\partial m}{\partial\alpha} D_{(\alpha)} + \frac{\partial m}{\partial\beta} D_{(\beta)} \right) \int_{\bf p} \frac{p^{\langle\mu\rangle}p^{\langle\nu\rangle}}{(p\cdot u)^2}\delta f_1 - \theta \, M\left( \frac{\partial M}{\partial\alpha} D_{(\alpha)} + \frac{\partial M}{\partial\beta} D_{(\beta)} \right) \int_{\bf q} \frac{p^{\langle\mu\rangle}p^{\langle\nu\rangle}}{(p\cdot u)^2}\delta f^+_1 \\
       &+ 2 m\left[ \frac{\partial m}{\partial\alpha} \nabla^{(\mu}\alpha + \frac{\partial m}{\partial\beta} \left( \frac{\rho}{\ped +\pres} \nabla^{(\mu}\alpha -\beta \, \dot u^{(\mu} \right) \right]\int_{\bf p}\frac{p^{\langle\nu\rangle)}}{(p\cdot u)} \delta f_1 \\
       &+ 2 M\left[ \frac{\partial M}{\partial\alpha} \nabla^{(\mu}\alpha + \frac{\partial M}{\partial\beta} \left( \frac{\rho}{\ped +\pres} \nabla^{(\mu}\alpha -\beta \, \dot u^{(\mu} \right) \right]\int_{\bf q}\frac{p^{\langle\nu\rangle)}}{(p\cdot u)} \delta f_1^+ \\
       &-\Delta^\mu_\alpha \Delta^\nu_\beta\nabla_\lambda\left[ \int_{\bf p} \frac{p^{\langle\lambda\rangle}p^{\langle\alpha\rangle}p^{\langle\beta\rangle}}{(p\cdot u)}\delta f_1 +\int_{\bf q} \frac{p^{\langle\lambda\rangle}p^{\langle\alpha\rangle}p^{\langle\beta\rangle}}{(p\cdot u)}\delta f^+_1  \right] \\
        & -\left( \sigma_{\lambda\rho} + \frac{1}{3}\, \theta \, \Delta_{\lambda\rho} \right) \left[ \int_{\bf p} \frac{p^{\langle\lambda\rangle}p^{\langle\rho\rangle}p^{\langle\mu\rangle}p^{\langle\nu\rangle}}{(p\cdot u)^2} \delta f_1 +\int_{\bf q} \frac{p^{\langle\lambda\rangle}p^{\langle\rho\rangle}p^{\langle\mu\rangle}p^{\langle\nu\rangle}}{(p\cdot u)^2} \delta f^+_1 \right] \\
        &+ \theta\, \Pi \, \Delta^{\mu\nu} - \frac{5}{3}\, \theta \, \pi^{\mu\nu} +2\, \Pi \,\left( \sigma^{\mu\nu} +\frac{1}{3} \, \theta \, \Delta^{\mu\nu} \right) - 2\sigma^{\lambda(\mu}\pi^{\nu)}_\lambda -2 \omega^{\lambda(\mu}\pi^{\nu)}_\lambda.
    \end{split}
\end{equation}

The $\delta f_1$ is obtained in the same way as its counterparts $\delta f^\pm$, and reads
\begin{equation}\label{delta_f_both}
    \begin{split}
    \delta f_1 \simeq & \frac{\taueq}{\kappa_b}\, \frac{\rho}{(\ped+\pres)} (p\cdot \nu) \, \fn +\frac{\taueq \, \beta}{2\eta} \,\frac{p\cdot \pi\cdot p}{(p\cdot u)} \, \fn \\
    & - \frac{\taueq \, \Pi}{\zeta}\left[ (p\cdot u) D_{(\beta)} -\frac{\beta(-p\cdot\Delta\cdot p)}{3(p\cdot u)} +\frac{\beta \, M}{(p\cdot u)} \left( \frac{\partial M}{\partial \alpha} D_{(\alpha)} + \frac{\partial M}{\partial \beta} D_{(\beta)} \right)\right]\fn \; .
    \end{split}
\end{equation}
Taking the traceless projection, one has the evolution of the shear pressure corrections $\pi^{\mu\nu}$
%Almost there, we've got the huge second-order formula Very long, but taking the traceless part one has the second-order evolution of $\pi^{\mu\nu}$ with all the second-order transport coefficients explicitly written (after some factorizations...). Making the contraction with $-\frac{1}{3}\Delta_{\mu\nu}$ one has the same for the bulk.
%The simplest is the shear, introducing the notation
%
%\begin{equation}
 %   {\cal O}^{\langle \mu\nu \rangle \cdots} = \Delta^{\mu\nu}_{\alpha\beta}{\cal O}^{\alpha\beta \cdots}=\frac{1}{2}\left( \Delta^\mu_\alpha \Delta^\nu_\beta + \Delta^\mu_\beta \Delta^\nu_\alpha -\frac{2}{3}\Delta^{\mu\nu}\Delta_{\alpha\beta}\right){\cal O}^{\alpha\beta\cdots},
%\end{equation}
%
\begin{align}\label{secondordershear}
\dot{\pi}^{\langle\mu\nu\rangle}+\frac{\pi^{\mu\nu}}{\taueq}=&\frac{2\, \eta}{\taueq} \, \sigma^{\mu\nu}-2\,\omega_{\lambda}^{\langle\mu}\pi^{\nu\rangle\lambda}+\tau_{\pi\pi}\,\sigma^{\langle\mu}_{\lambda}\pi^{\nu\rangle\lambda}+\delta_{\pi\pi}\,\pi^{\mu\nu}\theta+\lambda_{\pi\Pi}\Pi\sigma^{\mu\nu}+\tau_{\pi\nu}\,\dot{u}^{\langle\mu}\nu^{\nu\rangle}-\gamma_{\pi\nu}\, \nu^{\langle\mu}\nabla^{\nu\rangle}\alpha\nonumber\\
&~~~~~~~~~~~~~-\Delta^{\mu\nu}_{\alpha\beta}\nabla_\lambda \left[ l_{\pi\nu}\left( \Delta^{\lambda\alpha}\, \nu^\beta +\Delta^{\lambda\beta}\nu^\alpha\right) \right] .
\end{align}
Contracting with $-\Delta_{\mu\nu}/3$ one has the equation for the bulk $\Pi$ 
\begin{align}\label{2ndorderbulk}
    \dot{\Pi}+\frac{\Pi}{\taueq}=-\frac{\zeta}{\taueq}\, \theta \;+\;\delta_{\Pi\Pi}\,\theta \, \Pi \;+\; \lambda_{\Pi\pi}\,(\sigma:\pi) \;-\; \tau_{\Pi\nu}\, (\dot{u}\cdot\nu) \;+\; l_{\Pi\nu} \, (\partial\cdot\nu) \;+\; 
    n_{\Pi\nu}\,(\nu\cdot\nabla\alpha) +\frac{5}{3}\nabla\cdot\left(  l_{\pi\nu} \, \vphantom{\frac{}{}} \, \nu\right)~.
\end{align}
All the second-order transport coefficients for the pressure corrections are listed in Appendix~\ref{app_tc}, after the ones from the diffusion current.
%
%***************************
\section{Conclusion and Discussion}\label{sec:conclusions}
%
\iffalse
This work derived the transport coefficients in relativistic second-order dissipative hydrodynamics using an effective Boltzmann equation for a system of quasiparticle quarks, anti-quarks, and gluons. We incorporated temperature and chemical potential dependent masses, establishing a thermodynamically consistent framework. The framework describes second-order dissipative evolution equations, including corrections for shear and bulk viscous pressure, along with baryon charge diffusion.
%

Our work will offer the ability to integrate a realistic equation of state from kinetic theory. To explore this advantage, we will examine in the future the impact of our formulation on the viscous hydrodynamic evolution of strongly interacting matter created in one-dimensional, purely longitudinal, boost-invariant expansion following relativistic heavy-ion collisions. This paves the way for further investigation into the effects of this novel approach.
\fi

In this work, we have derived the equations of second-order viscous hydrodynamics which incorporates conserved baryonic current with a realistic equation of state in presence of non-vanishing chemical potential. We employed a quasiparticle model with an additional bag-like tensor degree of freedom as an intermediate approximation. This bag term is necessary to ensure both thermodynamic consistency at global equilibrium as well as the local conservation of charges (energy-momentum and baryon number). We have discussed in detail regarding the origin of such a model within quantum field theory and its link to the Wigner formalism. The method proposed in the present work is rather flexible and can be used with all the practical candidates for a realistic equation of state for nuclear matter.

%It cannot represent all the cases. Very extreme equations of state with large negative pressure compared to the energy (e.g. dark energy with $T^{\mu\nu}\propto g^{\mu\nu}$) are excluded. On the other hand, it can be used with all the practical candidates for a realistic equation of state for nuclear matter.

\bigskip
\bigskip

%***************************

{{\bf Acknowledgements:}}{ A.D. acknowledges discussions with Samapan Bhadury. A.J. acknowledges the kind hospitality of Jagiellonian University and IIT Gandhinagar where part of this work was carried out. This work was supported in part by the Polish National Science Centre Grants No. 2018/30/E/ST2/00432 (A.D.) and No. 2020/39/D/ST2/02054. (L.T. and R.R.).

\medskip 

%%%%%%%%%%%%%%%%%%%%%%%%%%%%%%%%%%%%%%%%%%%%%%%%%%%%%%%%%%%%%%%%%%%%%%%%%%%

\appendix

%***************************
\section{on-shell distribution functions}
%***************************
\label{sec:app_on-shell}

In this work, we have been using exclusively $f$, $f^q$, and $f^{\bar q}$ that depend exclusively on all four of the components of the vector $p^\mu$. The integrals in (\ref{def_int}) are on-shell and positive. It is not necessary to consider off-shell functions, that is with any $p^\mu p^\nu g_{\mu\nu}$, only their on-shell restriction will ever enter in the physical quantities. Many authors prefer to consider only such shell restrictions. In particular, if there is an unambiguous preferred reference frame, for instance, if the initial conditions are given at a fixed time in a specific frame. Without loss of generality, we will consider just one case, say $f$, the other cases are the same. If one wants to use the on-shell restriction of $f$ there is no need to make four-dimensional momentum integrals
\begin{equation}
    \frac{g}{(2\pi)^3}\int d^4 p \; 2\Theta(p_0) \delta(p^2 -m^2) \qquad \longrightarrow \qquad \frac{g}{(2\pi)^3}\int \frac{d^3p}{E_{\bf p}},
\end{equation}
with $E_{\bf p}$ clearly the on-shell, positive energy. It comes from the generic relation
\begin{equation}
    \delta (f(x)) \sum_{x_0} \frac{\delta(x-x_0)}{|f^\prime(x_0)|}, \qquad \forall \, x_0\, \mbox{ such that} \; f(x_0)=0.
\end{equation}
In all the polynomials in the $p^\mu$ components, such as the weighted integrals in $T^{\mu\nu}$ and $J^\mu$, the $p^0$ must be substituted by the on-shell energy, as usual. In order to be consistent then, the on-shell restriction of $f$ reads
\begin{equation}
    \bar f(x, {\bf p}) = \int dp_0 \; 2 \Theta(p_0) \, \delta (p^2 -m^2) \, p^0 f(x,p).
\end{equation}
The $p^0$ in the integral is the usual convention. It ensures that the physical dimensions of $f$ and $\bar f$ are the same and that the energy on the denominator of the three-momentum integrals is not counted twice.
On a similar note, it can be introduced the on-shell version of the four-momentum
\begin{equation}
    \bar p =(E_{\bf p},{\bf p}).
\end{equation}
For a generic Vlasov term in the evolution equation
\begin{equation}\label{Vlas_term}
    F_\mu \partial^\mu_{(p)} f ,
\end{equation}
the force term $F$ can depend on the position, momentum, or both, like in the case of the Lorentz force from the coupling with external electromagnetic fields in plasma physics. One can use the $\bar F$ notation here too, indicating that the energy in $F$ is replaced by the on-shell $E_{\bf p}$. One has then making use of Eq.~(\ref{delta_prime_prop})
\begin{equation}
    \begin{split}\label{on-shell_Vlasov}
        \bar F_i \;\frac{\partial \bar f}{\partial p_i} &= \bar F_i\int d p_0 \; 2 \Theta (p_0)\left[ g^{0i} \,\delta(p^2 - m^2) \, f \;+\; p^0 \, 2 \, p^i \, \delta^\prime(p^2 -m^2) f \;+\; p^0 \, \delta(p^2-m^2) \, \frac{\partial f} {\partial p_i} \right] \\
        &= \int d p_0 \; 2 \Theta (p_0)\left[ g^{0i} \,\delta(p^2 - m^2) \, f \, - \, g^{0i} \,\delta(p^2 - m^2) \, f \;-\; p^i \, \delta(p^2-m^2) \, \frac{\partial f} {\partial p_0}\;+\; p^0 \, \delta(p^2-m^2) \, \frac{\partial f} {\partial p_i} \right]F_i \\
        &= \int d p_0 \; 2 \Theta (p_0)\left[ -\; p^\mu \, F_\mu \, \delta(p^2-m^2) \, \frac{\partial f} {\partial p_0}\;+\; p^0 \, \delta(p^2-m^2) \, F_\mu \frac{\partial f} {\partial p_\mu} \right].
    \end{split}
\end{equation}
The last term on the right-hand side is just the Vlasov term~(\ref{Vlas_term}) integrated over $p_0$ according to the prescription used for $\bar f$ itself. The additional term is vanishing if the force term is perpendicular to the four-momentum $p\cdot F=0$. Without loss of generality, it is vanishing if it is perpendicular on-shell, because of the integration. This is the case of the Lorentz force $\propto F^{\mu\nu}p_\nu$ because of the antisymmetry of the Faraday tensor $F^{\mu\nu}$. In general, it does not go away. Definitely, not for an $F_\mu = \frac{1}{2}\partial_\mu m^2$ as in this work that does not depend on the momentum at all and cannot be orthogonal for all (on-shell positive) $p^\mu$. However, in this case, it is compensated by an equal and opposite sign term stemming from $\bar p\cdot \partial \bar f$. It is interesting to note that passages in~\ref{on-shell_Vlasov} are perfectly valid in a generic reference frame, not only on an inertial one. In fact, the compensation between the $g^{0i}$ terms is more important here, since $g^{0i}$ is not vanishing in a generic frame.

In this work, we consider, for convenience, only an inertial reference frame.  There is also a clash between the notation in hydrodynamics $\nabla_\mu$, the part of the gradient orthogonal to the four-velocity, and the $\nabla_\mu$ for the covariant derivatives, as opposed to the partial derivatives, in curvilinear frames. Since, however, the compensation of the term $p\cdot F$ in the right-hand side of~\ref{on-shell_Vlasov} from the on-shell $\bar p\cdot \partial \bar f$ happens, not only for $F_\mu=\frac{1}{2}\partial_\mu m^2$, but also for the apparent forces in curvilinear frames. Being them accelerated frames, or just the Milne coordinates that are used so often in heavy-ion applications. For completeness' sake, we give the final part of the proof in a generic frame.

Before that, it is necessary to write the Boltzmann-Vlasov equation, to see how to properly write the apparent forces. The Procedure in~(\ref{def_Vlas}) must be generalized. Considering the vector\footnote{Nonconserved, in general. For particles in relativistic kinetic theory it would be particle number flux, an observable. In the generalized sense it does not correspond to any observables, but it can be used to derive equations as it has been done in the main text.} $N^\mu$
\begin{equation}
    N^\mu=\frac{g}{(2\pi)^3} \int \frac{d^4 p}{\sqrt{-g}} \; 2 \theta(p_0) \, \delta(p^2 -m^2) \, p^\mu \, f.
\end{equation}
The square root $\sqrt{-g}$ is the standard short-hand notation for the root of the modulus of the determinant of $g_{\mu\nu}$. It is there, as usual, to ensure the absolute value of the Jacobian of the transformation from the Cartesian coordinates is taken into account properly. The choice of considering the covariant $p_{\mu}$ as variables of integration instead of the contravariant $p^\mu$ is not mandatory. It simplifies some passages, but the same arguments can be used in the contravariant case reaching similar results\footnote{Not identical though. The equations are not tensorial after all. The function $f(x, p)$ is not linear in $p$, so not a contraction with $p$ seen as a (covariant or contravariant) field. In a generic frame the $p^\mu = g^{\mu\nu} p_\nu$ are rather different from the $p_\nu$, they do not change by just a sing (if any) as in the inertial frames.}.

The divergence takes a term in the connection coefficients, as usual for covariant derivatives, which must be taken into account
\begin{equation}
    \begin{split}\label{Vlas_non-inert}
       \partial_\mu N^\mu + \Gamma^\mu_{\mu\nu} N^\nu &= \frac{g}{(2\pi)^3}\int\frac{d^4 p}{\sqrt{-g}} \; 2\Theta(p_0)\left[ -\frac{1}{2}g^{\alpha\beta}\partial_\mu \, g_{\alpha\beta} \, \delta(p^2 -m^2) p^\mu f \;+\: \partial_\mu(p^2 -m^2)\, \delta^\prime(p^2 -m^2) \, p^\mu f \right. \\
       &\left. \qquad \qquad\qquad  \;+\; \delta(p^2 -m^2)\partial_\mu \, g^{\mu\nu}p_\nu f \;+\:\delta(p^2-m^2) \, p\cdot\partial f \;+\; \delta(p^2-m^2) \, \Gamma^\mu_{\mu\nu}p^\nu \, f \vphantom{\frac{1}{2}}\right]\\
       &= \frac{g}{(2\pi)^3}\int\frac{d^4 p}{\sqrt{-g}} \; 2\Theta(p_0) \delta(p^2 -m^2)\left\{\left[ -\frac{1}{2}g^{\alpha\beta} p\cdot\partial \, g_{\alpha\beta} \;-\partial_\mu \, g^{\mu\nu}p_\nu \; + \partial_\mu \, g^{\mu\nu}p_\nu \;+ \; \Gamma^\mu_{\mu\nu} p^\nu\right] f \right. \\
       & \qquad \qquad \qquad \qquad \qquad\left. \;+\; p\cdot \partial f \;- \frac{1}{2}\partial_\mu(p^2-m^2)\partial^\mu_{(p)}f \right\}\\
       &= \frac{g}{(2\pi)^3}\int\frac{d^4 p}{\sqrt{-g}} \; 2\Theta(p_0) \delta(p^2 -m^2)\left[ p\cdot\partial f \;-\; \frac{1}{2} \partial_\mu\left( \vphantom{\tfrac{1}{2}} p^2 -m^2 \right)\partial^\mu_{(p)}f \right].
    \end{split}
\end{equation}
The last passage might be not immediate, but it can be seen from the compatibility of the covariant derivatives with $g_{\mu\nu}$
\begin{equation}\label{comp_g}
    \begin{split}
        &\partial_\alpha \, g_{\mu\nu} = \Gamma^\beta_{\mu\alpha}\, g_{\beta\nu} \;+\; \Gamma^\beta_{\nu\alpha}\, g_{\beta\mu} \\
        & \Rightarrow -\frac{1}{2}g^{\alpha\beta} \left(\vphantom{\frac{}{}} \partial_\nu \, g_{\alpha\beta} = \Gamma^\rho_{\nu\alpha}\, g_{\rho\beta} \;+\; \Gamma^\rho_{\nu\beta}\, g_{\rho\alpha}\right) = -\frac{1}{2}\left( \vphantom{\frac{}{}} \Gamma^\alpha_{\nu\alpha} + \Gamma^\beta_{\nu\beta} \right) = -\Gamma^\mu_{\mu\nu}.
    \end{split}
\end{equation}
The Vlasov term in a generic frame is thus a rather simple extension of the Cartesian one
\begin{equation}
    p\cdot\partial f -\frac{1}{2}\partial_{\mu}\left( \vphantom{\frac{}{}}p^2-m^2 \right) \; \partial^\mu_{(p)} f,
\end{equation}
the difference being that the gradient of the (inverse of the) metric $\partial_\mu g^{\alpha\beta}$ is not vanishing in general. As expected, it coincides with the regular formula in Cartesian coordinates. The $\bar p\cdot \bar f$ term is indeed providing the needed term$-\frac{1}{2}p\cdot\partial(p^2-m^2)\; f$ that compensates the $-p\cdot F$ in the right hand side of~\ref{on-shell_Vlasov}. Indeed
\begin{equation}
    \begin{split}
        \bar p\cdot \partial \bar f &= \int dp_0 \; 2\Theta(p_0)\left[ p\cdot\partial \, g^{0\nu} p_\nu \, \delta(p^2-m^2) \, f \;+\; p^0 \, \bar p\cdot\partial\left( p^2-m^2 \vphantom{\frac{}{}} \right) \, \delta^\prime(p^2 -m^2) f \;+\; p^0 \delta(p^2 -m^2) \, p\cdot \partial f \right] \\
        &= \int dp_0 \; 2\Theta(p_0)\, \delta(p^2-m^2)\left\{ \left[ p\cdot\partial \, g^{0\nu} p_\nu -\vphantom{\frac{1}{2}}p\cdot\partial \, g^{0\mu} p_\mu \right] f \;-\; \frac{1}{2} p\cdot \partial \left( p^2-m^2 \vphantom{\frac{}{}} \right) \, \, \frac{\partial f}{\partial p_0} \;+\; p\cdot \partial f \right\} \\
        &= \int dp_0 \; 2\Theta(p_0)\, \delta(p^2-m^2) \left[ \;-\; \frac{1}{2} p\cdot \partial \left( p^2-m^2 \vphantom{\frac{}{}} \right) \, \, \frac{\partial f}{\partial p_0} \;+\; p\cdot \partial f \right].
    \end{split}
\end{equation}
In order to avoid confusion, it is convenient to remember that a naive substitution of the off-shell momentum $p$ with the on-shell momentum $\bar p$
in the force term to obtain $\bar F$. This is because the gradient $\partial_\mu \bar p^2$ contains, on top of the gradients of $g^{\alpha \beta}$ that should be there, also the gradients of the on-shell\footnote{In a generic frame the on-shell $p_0$ has a more complicated formula than in inertial frames. It does depend both on the metric, or, better, its inverse, and the mass $p_0 = \frac{1}{g^{00}}\left[ -g^{0i}p_i +\sqrt{(g^{0i}p_i)^2+g^{00}(m^2 - g^{ij}p_ip_j)} \right]$, see for instance~\cite{DEBBASCH20091079}. Each of them is a source of spacetime dependence.} $p_0$ that shouldn't be there. It is convenient to pull the $ p$ out of the derivatives to have a safe substitution. This can be done by writing the term with the aid of the connection coefficients too, see also Ref.~\cite{DEBBASCH20091079}
\begin{equation}
    -\frac{1}{2}\, p_\alpha p_\beta \, \partial_\mu g^{\alpha\beta} =\frac{1}{2}\, p_\alpha p_\beta\left( \vphantom{\frac{}{}} \Gamma^\alpha_{\mu \rho}g^{\rho\beta} + \Gamma^\beta_{\mu\rho}g^{\rho\alpha} \right) = \Gamma^\alpha_{\mu\beta}\;p_\alpha p^\beta,
\end{equation}
therefore
\begin{equation}
    -\frac{1}{2}\partial_\mu \left( \vphantom{\frac{}{}} p^2-m^2\right) = p_\nu\, \Gamma^\nu_{\mu\rho}\, p^\rho + m\, \partial_\mu m.
\end{equation}
On the right-hand side, it is safe to substitute the $p$ with the $\bar p$ since they are not under any derivation. In the end, the link between the on-shell and off-shell formulation of the Botlmann-Vlasov equation is
\begin{equation}
\begin{split}
    \bar p\cdot \partial \bar f + \left( \vphantom{\frac{}{}} \bar p_\rho \, \Gamma^\rho_{i\nu} \, \bar p^\nu \; + \; m\, \partial_i m  \right) \, \partial^i_{(p)} f &= \int dp_0 \; 2\Theta(p_0)\, \delta(p^2 - m^2) \, p^0 \left[ p\cdot \partial f \; +\; \left( \vphantom{\frac{}{}} p_\rho \, \Gamma^\rho_{\mu\nu} \, p^\nu \; + \; m\, \partial_\mu m  \right) \, \partial^\mu_{(p)} f \right]\\
    &=\int dp_0 \; 2\Theta(p_0)\, \delta(p^2 - m^2) \, p^0 \left[ p\cdot \partial f \; -\; \frac{1}{2}\partial_\mu\left( \vphantom{\frac{}{}} p^2 -m^2 \right) \, \partial^\mu_{(p)}f \right].
\end{split}
\end{equation}
That is, one is the positive, on-shell integral of the other, using the very same prescription employed for $\bar f$. Needless to say, the generalized collisional kernel, whether it is of the relaxation type or not, must be made on-shell using the same integral in the on-shell picture.

%***************************
\section{Transport coefficients}
%***************************
\label{app_tc}

The transport coefficients obtained in this work have rather bulky formulas. We report them here, in order to avoid such cumbersome notation in the main text.

The second-order transport coefficients appearing in the diffusion current equation~\eqref{diffusion_compact} read
\begin{align}
    &\tau_{\nu \Pi}=\frac{\rho}{(\ped+\pres)} \;-\;\frac{q \,\taueq \, \beta\, M}{\zeta}\, \frac{\partial M}{\partial\beta}\left[q\, D_{(\alpha)}\, I^{+}_{-1,0}-D_{(\beta)}\, I_{0,0}^{-} + \beta\,  I^{-}_{0,1}-\beta M\left(\frac{\partial M}{\partial\alpha}D_{(\alpha)}+\frac{\partial M}{\partial \beta}D_{(\beta)}\right)I^{-}_{-2,0}\right] ,\\
    &c_{\pi\Pi} = \frac{\rho}{(\ped+\pres)}, \\
    &\delta_{\nu\nu}=-\frac{q\, \taueq\, M}{\kappa_{b}}\left(\frac{\partial M}{\partial\alpha}D_{(\alpha)}+\frac{\partial M}{\partial \beta}D_{(\beta)}\right)\left[q I_{-1,1}^{+}-\frac{\rho}{(\ped+\pres)}I_{0,1}^{-}\right]+\frac{5\,q\,\taueq}{3\,\kappa_{b}}\left[ q \,I_{1,2}^{+}-\frac{\rho}{(\ped+\pres)}I_{2,2}^{-}\right] -\frac{4}{3},\\
    &c_{\nu\Pi} = \frac{q \,\taueq \, M}{\zeta}\left( \frac{\partial M}{\partial\alpha} \,+\, \frac{\rho}{(\ped+\pres)} \frac{\partial M}{\partial\beta} \right)\left[q\, D_{(\alpha)}\, I^{+}_{-1,0}-D_{(\beta)}\, I_{0,0}^{-} + \beta\,  I^{-}_{0,1}-\beta M\left(\frac{\partial M}{\partial\alpha}D_{(\alpha)}+\frac{\partial M}{\partial \beta}D_{(\beta)}\right)I^{-}_{-2,0}\right],\\
    &l_{\nu\Pi}=\frac{q\, \taueq}{\zeta}\left[q\, D_{(\alpha)}\, I^{+}_{1,1} - D_{(\beta)}\, I_{2,1}^{-}+\frac{5}{3}\, \beta \, I^{-}_{2,2}-\beta M\left(\frac{\partial M}{\partial\alpha}D_{(\alpha)}+\frac{\partial M}{\partial \beta}D_{(\beta)}\right)I^{-}_{0,1}\right],\\
    &l_{\nu\pi}=\frac{q\, \taueq \, \beta}{\eta} \, I_{2,2}^{-}\\
    &\lambda_{\nu\nu}=\frac{2\,q \,\taueq}{\kappa_{b}}\left[q I_{1,2}^{+}-\frac{\rho}{(\ped+\pres)}I_{2,2}^{-}\right] -1.
\end{align}
The ones in the shear pressure corrections evolution equation\eqref{secondordershear} are  
\begin{align}\label{2ndordersheartransport}
&\tau_{\pi\pi}=\frac{\taueq}{\eta}4\beta\left(I^{+}_{3,3}+I_{3,3}\right)-2,\\
&\delta_{\pi\pi}=\frac{\taueq}{\eta}\left[\frac{7}{3}\beta(I^{+}_{3,3}+I_{3,3})-M \beta\left(\frac{\partial M}{\partial \alpha}D_{(\alpha)}+\frac{\partial M}{\partial\beta}D_{(\beta)}\right) I^{+}_{1,2}-m\beta\left(\frac{\partial m}{\partial \alpha}D_{(\alpha)}+\frac{\partial m}{\partial\beta}D_{(\beta)}\right)I_{1,2} \right]-\frac{5}{3},\\
&\lambda_{\pi\Pi}=\frac{2\, \taueq}{\zeta}\left[ \left(I_{3,2}+I^+_{3,2} \vphantom{\frac{}{}}\right)D_{(\beta)}-qI^-_{2,2}D_{(\alpha)} +  M\beta\left( \frac{\partial M}{\partial \alpha} D_{(\alpha)} + \frac{\partial M}{\partial \beta} D_{(\beta)} \right)I^+_{1,2} \right. \\ \nonumber
& ~~~~~~~~~~~~~~~~~~~~~~~~~\left. + m\beta\left( \frac{\partial m}{\partial \alpha} D_{(\alpha)} + \frac{\partial m}{\partial \beta} D_{(\beta)} \right)I_{1,2}\right]-\frac{14\, \taueq \, \beta}{3\, \zeta}\left(I^{+}_{3,3}+I_{3,3}\right)+2,\\
&\tau_{\pi\nu}=2\,\beta\,\frac{\taueq}{\kappa_{b}}\left[ M\, \frac{\partial M}{\partial\beta}\left(-q\,I^{-}_{0,1}+\frac{\rho}{\ped+\pres}\, I^{+}_{1,1}\right)+m\, \frac{\partial m}{\partial\beta} \, \frac{\rho}{\ped+\pres}\, I_{1,1}\right],\\
&\gamma_{\pi\nu}=2\, \frac{\taueq}{\kappa_{b}} \left[M  \left(\frac{\partial M}{\partial \alpha}+\frac{\rho}{\ped+\pres}\frac{\partial M}{\partial\beta}\right)\left(-q\, I^{-}_{0,1}+\frac{\rho}{\ped+\pres}\, I^{+}_{1,1}\right)+m\left(\frac{\partial m}{\partial \alpha}+\frac{\rho}{\ped+\pres}\, \frac{\partial m}{\partial\beta}\right)\frac{\rho}{\ped+\pres}\, I_{1,1}\right],\\
&l_{\pi\nu}=\frac{\taueq}{\kappa_{b}}\left[\frac{\rho}{\ped+\pres}\left(I^+_{3,2}+ I_{3,2}\right)-q\, I^{-}_{2,2}\right].\label{sheartransportend}
\end{align}
The remaining ones, in the bulk equation~\eqref{2ndorderbulk}, read
\begin{align}
    &\delta_{\Pi\Pi}=M\left( \frac{\partial M}{\partial\alpha }D_{(\alpha)} + \frac{\partial M}{\partial\beta}D_{(\beta)} \right)\frac{\taueq}{\zeta}\left[I^+_{1,1}D_{(\beta)} - q\, I^{-}_{0,1}D_{(\alpha)}-\frac{5\beta}{3}I^+_{1,2} +\beta M \left(  \frac{\partial M}{\partial \alpha} D_{(\alpha)} + \frac{\partial M}{\partial \beta} D_{(\beta)}  \right)I^+_{-1,1}\right]\nonumber\\
    &~~~~~~~~~-\, m\left( \frac{\partial m}{\partial\alpha }D_{(\alpha)} + \frac{\partial m}{\partial\beta}D_{(\beta)} \right)  \,\frac{\taueq}{\zeta}\left[I_{1,1}D_{(\beta)} -\frac{5\beta}{3}I_{1,2} +\beta m \left(  \frac{\partial m}{\partial \alpha} D_{(\alpha)} + \frac{\partial m}{\partial \beta} D_{(\beta)}  \right)I_{-1,1}\right]\nonumber\\
    &~~~~~~~~~+(I_{3,3}^{+}+I_{3,3})\frac{35}{9}\frac{\beta\taueq}{\zeta}-\frac{5}{3}-\frac{5\taueq}{3\zeta}\bigg[\left(I_{3,2}+I^+_{3,2} \vphantom{\frac{}{}}\right)D_{(\beta)}-{q\, I^-_{2,2}D_{(\alpha)}}+M{\beta}\left( \frac{\partial M}{\partial \alpha} D_{(\alpha)} + \frac{\partial m}{\partial \beta} D_{(\beta)} \right){I^+_{1,2}}\nonumber\\
    &~~~~~~~~~+ m{\beta}\left( \frac{\partial m}{\partial \alpha} D_{(\alpha)} + \frac{\partial m}{\partial \beta} D_{(\beta)} \right){I_{1,2}}\bigg] \\\nonumber
    &~~~~~~~~~-m\left( \frac{\partial m}{\partial\alpha }D_{(\alpha)} + \frac{\partial m}{\partial\beta}D_{(\beta)}  \right) \left[ D_{(\beta)} I_{1,0}-\beta \, I_{1,1} -\beta m\left( \frac{\partial m}{\partial\alpha }D_{(\alpha)} + \frac{\partial m}{\partial\beta}D_{(\beta)}  \right) I_{-1,0}  \right] \\\nonumber
    &~~~~~~~~~-M\left( \frac{\partial M}{\partial\alpha }D_{(\alpha)} + \frac{\partial M}{\partial\beta}D_{(\beta)}  \right) \left[ D_{(\beta)} I_{1,0}^+ -q\, D_{(\alpha)} I^-_{0,0}-\beta \, I_{1,1} -\beta m\left( \frac{\partial m}{\partial\alpha }D_{(\alpha)} + \frac{\partial m}{\partial\beta}D_{(\beta)}  \right) I_{-1,0}  \right] \\ \nonumber
    &~~~~~~~~~~-\frac{(\ped+\pres)\, \rho_\beta}{\beta\, {\cal J}} -\frac{\rho\, \rho_\alpha}{\beta \, {\cal J}}.
    \label{2ndorderbulktransport}\\
    &\lambda_{\Pi\pi}=\frac{2}{3} -\frac{7\, \taueq\, \beta}{3\, \eta}\left( I^{+}_{3,3}+I_{3,3} \right) + \frac{(\ped+\pres)\, \rho_\beta}{\beta\, {\cal J}} + \frac{\rho\, \rho_\alpha}{\beta \, {\cal J}}\\
    &\tau_{\Pi\nu}=\frac{2 \,\beta\, \taueq}{3\, \kappa_b}\left[ M  \, \frac{\partial M}{\partial \beta}\left(\frac{\rho}{\ped+\pres}\, I^{+}_{1,1} -q\, I^{-}_{0,1}\right) + m  \, \frac{\partial m}{\partial\beta} \left( \frac{\rho}{\ped+\pres}\right) I_{1,1} \right].\\
    &l_{\Pi\nu}= \frac{(\ped+\pres)\, \ped_\beta}{\beta\, {\cal J}} + \frac{\rho\, \ped_\alpha}{\beta \, {\cal J}}.\\
    &n_{\Pi\nu}=\frac{2 \, \taueq}{3\, \kappa_b}\left[ M  \, \frac{\partial M}{\partial \alpha}\left(\frac{\rho}{\ped+\pres}\, I^{+}_{1,1} -q\, I^{-}_{0,1}\right) + m \, \frac{\partial m}{\partial\alpha} \left( \frac{\rho}{\ped+\pres}\right) I_{1,1} \right] +\frac{\rho \, \tau_{\Pi\nu}}{\beta(\ped +\pres)}~.\label{bulktransportend}
\end{align}

\bibliography{ref}
\end{document}